%% file: main.tex
\documentclass[]{style/iromlab}

\input{preamble/notation}

\usepackage[utf8]{inputenc}
\usepackage{multicol}
\usepackage{amsmath, amsfonts, amssymb, amsthm}
\usepackage{enumerate}
\usepackage[inline]{enumitem}
\usepackage{mathtools}
\usepackage{graphicx}
\usepackage{longtable,tabularx}
\usepackage{placeins} 
\usepackage{float}
\usepackage{multirow}
\usepackage{bbm}
\usepackage{threeparttable}
\usepackage{balance}
\usepackage[ruled,algo2e]{algorithm2e}
\usepackage{algpseudocode}
\usepackage{adjustbox}
\usepackage{booktabs}
\usepackage{bm}
\usepackage{etoolbox}
\usepackage{microtype}
\usepackage[title]{appendix}
\usepackage{units}
\usepackage{cleveref}
\usepackage{xspace}
\usepackage{tcolorbox}
\usepackage{caption}
\usepackage{tikz}
\usepackage{pifont}

\definecolor{claude_color}{HTML}{F89E62}
\definecolor{deepseek_color}{HTML}{78B6E8}
\definecolor{o3_mini_color}{HTML}{6CD5A1}
\definecolor{red_color}{HTML}{E13B55}
\definecolor{prompt_color}{HTML}{71502B}
\definecolor{wine}{RGB}{204, 0, 102}
\definecolor{survey}{RGB}{255, 207, 88}

\tcbset{
  promptstyle/.style={
    colback=gray!10,
    colframe=gray!60,
    boxrule=0.5pt,
    arc=2pt,
    outer arc=2pt,
    left=4pt,
    right=4pt,
    top=4pt,
    bottom=4pt,
    fonttitle=\bfseries,
    sharp corners=south,
  }
}

\tcbset{
  promptstyle_prompt_iuq/.style={
    colback=prompt_color!10,
    colframe=prompt_color!60,
    coltitle=white,
    boxrule=1.0pt,
    arc=2pt,
    outer arc=2pt,
    left=4pt,
    right=4pt,
    top=4pt,
    bottom=4pt,
    fonttitle=\bfseries,
  }
}

\tcbset{
  promptstyle_claude/.style={
    colback=claude_color!10,
    colframe=claude_color!60,
    coltitle=black,
    boxrule=1.0pt,
    arc=2pt,
    outer arc=2pt,
    left=4pt,
    right=4pt,
    top=4pt,
    bottom=4pt,
    fonttitle=\bfseries,
  }
}

\tcbset{
  promptstyle_deepseek/.style={
    colback=deepseek_color!10,
    colframe=deepseek_color!60,
    coltitle=black,
    boxrule=1.0pt,
    arc=2pt,
    outer arc=2pt,
    left=4pt,
    right=4pt,
    top=4pt,
    bottom=4pt,
    fonttitle=\bfseries,
  }
}

\tcbset{
  promptstyle_o3_mini/.style={
    colback=o3_mini_color!10,
    colframe=o3_mini_color!60,
    coltitle=black,
    boxrule=1.0pt,
    arc=2pt,
    outer arc=2pt,
    left=4pt,
    right=4pt,
    top=4pt,
    bottom=4pt,
    fonttitle=\bfseries,
  }
}

\usetikzlibrary{shapes.geometric, arrows}

\tikzstyle{rect} = [rectangle, 
rounded corners, 
minimum width=7cm,
minimum height=1cm,
text centered,
draw=black,
fill=red!30
]

\tikzstyle{rect_wrap} = [rectangle, 
rounded corners, 
minimum width=3cm,
minimum height=1cm,
text width=4cm,
text centered,
draw=black,
fill=red!30
]

\tikzstyle{rect_medium} = [rectangle, 
rounded corners, 
minimum width=3cm,
minimum height=1cm,
text width=5cm,
text centered,
draw=black,
fill=red!30
]

\tikzstyle{rect_large} = [rectangle, 
rounded corners, 
minimum width=5cm,
minimum height=1cm,
text width=5cm,
text centered,
draw=black,
fill=red!30
]

\tikzstyle{rect_larger} = [rectangle, 
rounded corners, 
minimum width=7cm,
minimum height=1cm,
text width=7cm,
text centered,
draw=black,
fill=red!30
]

\definecolor{row0_color}{RGB}{121, 181, 233}
\definecolor{row1_color}{RGB}{251, 172, 181}
\definecolor{row2_color}{RGB}{162, 214, 108}
\definecolor{row3_color}{RGB}{195, 163, 235}

\tikzstyle{arrow} = [thick, -, >=stealth]

\newbool{extended}
\setbool{extended}{false}

\makeatletter
\newcommand{\longdash}[1][2em]{%
  \makebox[#1]{$\m@th\smash-\mkern-7mu\cleaders\hbox{$\mkern-2mu\smash-\mkern-2mu$}\hfill\mkern-7mu\smash-$}}
\makeatother
\newcommand{\omitskip}{\kern-\arraycolsep}

\newcommand{\cmark}{\ding{51}}
\newcommand{\xmark}{\ding{55}}

\author[1*]{Zhiting Mei}
\author[1*]{Tenny Yin}
\author[1*]{Ola Shorinwa}
\author[1]{Apurva Badithela}
\author[1]{Zhonghe Zheng}
\author[2]{Joseph Bruno}
\author[1]{Madison Bland}
\author[1]{Lihan Zha}
\author[1]{Asher Hancock}
\author[1]{Jaime Fernández Fisac}
\author[2]{Philip Dames}
\author[1]{Anirudha Majumdar}

\affiliation[1]{Princeton University}
\affiliation[2]{Temple University}
\contribution[*]{Equal contribution.}

\begin{document}

\title{
\LARGE Robotic Video World Models:
{\LARGE A Survey of Applications, Research Challenges, Future Directions
}
}

\abstract{
    \input{sections/front_matter/abstract}
}

\keywords{
Video Generation Models, World Modeling, Visual Planning, Policy Learning and Evaluation \\[3ex]
}

\bibpage{
https://github.com/irom-princeton/awesome-robotics-video-world-model-papers %
}
{
github.com/irom-princeton/awesome-robotics-video-world-model-papers  %
}

\maketitle

\input{sections/front_matter/introduction}

\input{sections/background/background}

\input{sections/applications/applications}

\input{sections/evaluation/evaluation}

\input{sections/challenges_and_future_work/challenges_and_future_work}

\input{sections/back_matter/conclusion}

\input{sections/back_matter/acknowledgments}

\bibliographystyle{unsrtnat}
\bibliography{references.bib}

\end{document}

%% file: preamble/notation.tex
\DeclareDocumentEnvironment{example}{}{\noindent\textbf{Running example:}\itshape}{}

\usepackage{ifthen}
\newboolean{include-notes}
\newboolean{include-new}
\newboolean{include-remove}
\setboolean{include-notes}{true}
\setboolean{include-new}{false}
\setboolean{include-remove}{false}

\usepackage[dvipsnames]{xcolor}
\usepackage[normalem]{ulem}
\newcommand{\justin}[1]{\ifthenelse{\boolean{include-notes}}{\textcolor{orange}{\textbf{Jaime:} #1}}{}}

\newcommand{\princeton}[1]{\ifthenelse{\boolean{include-notes}}{\textcolor{orange}{#1}}{}}

\newcommand{\p}[1]{\smallskip \noindent \textbf{{#1}.}}

\providecommand{\onl}{\\[-0.7ex]}

%% file: sections/front_matter/abstract.tex
Video world models have emerged as promising candidates for high-fidelity world models, offering the potential to synthesize high-quality videos capturing fine-grained interactions between agents and their environments conditioned on multi-modal user inputs. Their impressive capabilities address many of the long-standing challenges faced by physics-based simulators, driving broad adoption in many problem domains, e.g., robotics.
For example, video models can generate photorealistic, physically consistent deformable-body simulation without making prohibitive simplifying assumptions, which is a major bottleneck in physics-based simulation.
Moreover, video models can serve as foundation world models that capture the dynamics of the world in a fine-grained and expressive way. They thus overcome the limited expressiveness of language-only abstractions in describing intricate physical interactions.
However, despite their potential, video models still generate physics-violating future predictions, often manifesting as hallucinations.
In this survey, we provide a review of video models and their applications as embodied world models in robotics, including efficient data generation and policy learning, dynamics and rewards modeling in reinforcement learning, policy evaluation, and visual planning. Further, we highlight important challenges hindering the trustworthy integration of video models, such as poor instruction following, hallucinations like violations of physics, and unsafe content generation, in addition to fundamental limitations such as significant data curation, training, and inference costs. We present potential future directions to address these open research challenges to motivate research and ultimately facilitate broader applications, especially in safety-critical settings. We provide a curated bibliography at \url{https://github.com/irom-princeton/awesome-robotics-video-world-model-papers}.

%% file: sections/front_matter/introduction.tex
\begin{figure}[H]
    \vspace{-0.5ex}
    \centering
    \includegraphics[width=\linewidth]{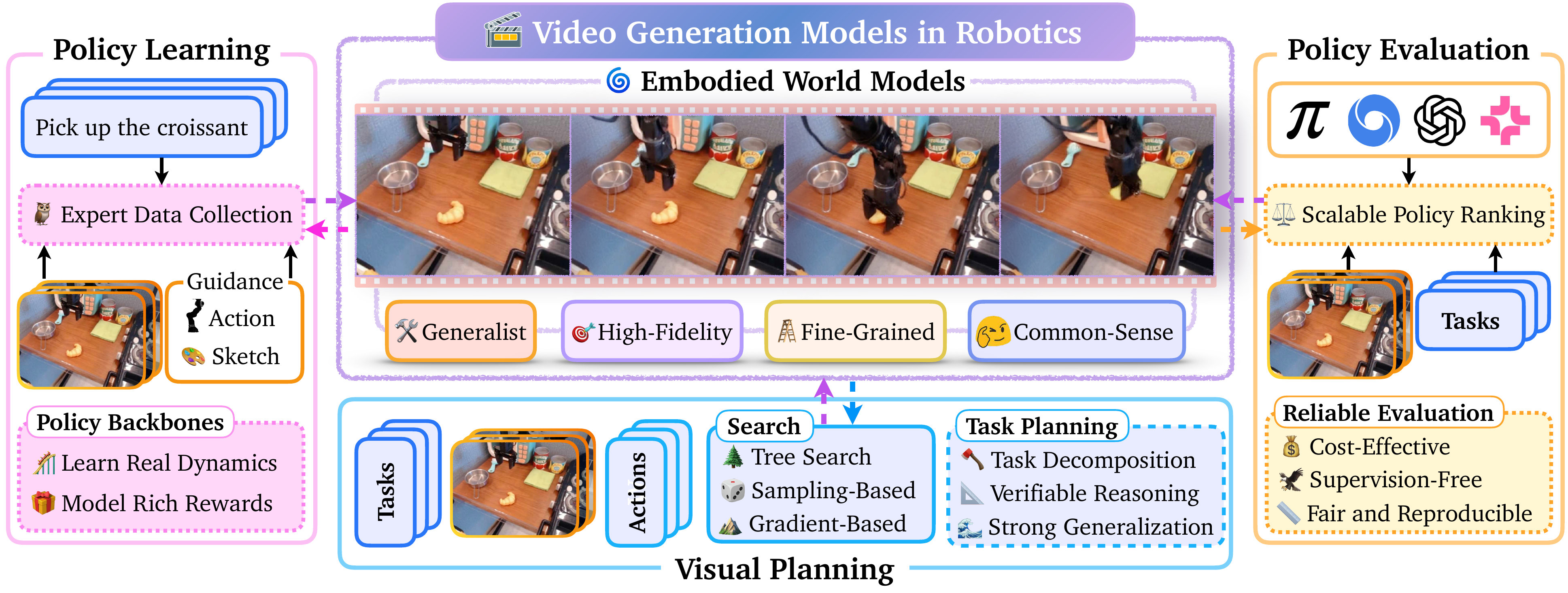}
    \caption{\textbf{Overview.} As embodied world models, video models offer immense potential in high-fidelity future prediction, overcoming some of the most notable challenges faced by classical physics-based simulators. These capabilities drive state-of-the-art advances in generalist robot policy learning, policy evaluation, and visual planning.
    }
    \label{fig:banner}
    \vspace{-0.5ex}
\end{figure}

\section{Introduction}
\label{sec:intro}
Recent breakthroughs in generative modeling, such as diffusion and flow-matching~\citep{ho2020denoising, lipman2023flowmatchinggenerativemodeling}, have enabled high-fidelity controllable video synthesis conditioned on user inputs such as text prompts, robot actions, and video frames. By training on internet-scale data, state-of-the-art (SOTA) video models learn to create rich videos that capture aesthetic and dynamic (motion) effects (e.g., with cinematic lighting, camera motion, and physical interaction between agents), driving widespread applications in video editing and content creation~\citep{li2025diffueraser, KuaishouKlingAI2024, RunwayGen3, LumaDreamMachine2024, brooks2024video, deepmind2025genie3, deepmind_veo3_techreport, wan2025wan, agarwal2025cosmos}.
As a result, these models have been increasingly integrated into robotics, e.g., in robot data generation, visual planning, policy learning, and policy evaluation~\citep{jang2025dreamgen, li2025novaflow, du2023videolanguageplanning, gao2024flip, quevedo2025worldgymworldmodelenvironment, guo2025ctrl}, drawing upon their impressive zero-shot generalization capabilities~\citep{wiedemer2025videomodelszeroshotlearners}.
In this survey, we provide a review of video models, highlighting their capabilities, applications in robotics, and limitations.

The advent of large vision-language models (VLMs/LLMs)~\citep{guo2025deepseek, OpenAI_GPT5_2025, Hassabis_Kavukcuoglu_Gemini3} has significantly transformed the state of the art in many problem domains (e.g., natural language processing, computer vision, robotics, etc.) due to their remarkable language generation and commonsense reasoning capabilities. Through pretraining on internet-scale data, LLMs learn broad foundational knowledge required to solve a wide range of tasks, which has resulted in their emergence as impressive zero-shot AI models. For example, VLMs function as the backbones of SOTA vision-language-action robot policies~\citep{zitkovich2023rt, intelligence2025pi_, team2025gemini}, which enable robots to perform diverse tasks with a single (unified) policy in place of multiple task-specific policies.
Despite these impressive capabilities, language models face some important limitations. First, language-only abstractions lack the expressiveness required to efficiently capture the intricate interaction processes inherent in the physical world. For example, consider the task of creating a concise fine-grained description of the contact interactions between a robot's gripper and a target deformable object, e.g., a cloth. Such a task is immensely challenging due to the high density of information and the limited capacity of language.
Second, language-centric modeling fails to accurately model the spatial and temporal dependencies between real-world phenomena (events) that are crucial to comprehensively understanding the physical world.
Although physics-based simulators offer visual world modeling capabilities, their practical effectiveness is limited by a number of fundamental challenges. Particularly, physics-based simulation generally requires restrictive simplifying assumptions that hinder the visual and physical fidelity of their predictions. For example, physics engines require expensive asset curation procedures to minimize the sim-to-real gap and struggle to accurately simulate deformable bodies with complex morphology and dynamics.
Video generation models offer a path towards overcoming these limitations by serving as photorealistic, physically consistent spatiotemporal models of the world. These capabilities, alongside their long-term potential, have catalyzed the adoption of video models across various robotics applications.

In robotics, video models have been increasingly adopted as embodied world models~\citep{huang2025vid2world, huang2025enerverse, liao2025genie, bruce2024genie}. 
High-fidelity 
world modeling establishes a trustworthy foundation for efficient evaluation of robot policies~\citep{guo2025ctrl, 1x_world, quevedo2025worldgymworldmodelenvironment, veorobotics2025}. Traditionally, \emph{policy evaluation} has necessitated setting up real-world robot stations for online policy roll-outs, which is notably expensive given the pertinent hardware and labor cost. Video models can circumvent this bottleneck without compromising the reliability of evaluation results.
Additionally, video models can provide high-fidelity dynamics and rewards predictions for training robot policies using reinforcement learning~\citep{hafner2025training, xiao2025world, escontrela2023video, huang2024diffusion}, while also enabling \emph{cost-effective robot data generation} in imitation learning. While data scaling has proven to be a critical recipe for advancing the performance of SOTA robot policies~\citep{GeneralistAI_GEN0_2025, intelligence2025pi_}, collecting expert demonstration data is incredibly expensive, posing a significant limitation. Video models can facilitate scalable data generation without relying on human supervision, addressing this pressing challenge. 
The expert demonstrations from video models can also be 
directly applied to a robot through motion retargeting~\citep{ajay2023compositional, xu2025vilp, ko2023learning}. 
Beyond generating successful task demonstrations, video models can synthesize failure video trajectories that endow robot policies with corrective behaviors for more robustness. 
Further, the generated data can be optimized through \emph{visual planning} to compute more optimal robot trajectories~\citep{ebert2018visual, gao2024flip, bu2024closed, du2023videolanguageplanning, chen2025largevideoplannerenables, pai2025mimic}.

Although video models provide valuable capabilities, their integration into robotics faces notable limitations. Like LLMs, video models tend to hallucinate, generating videos that are physically unrealistic, e.g., with objects appearing/disappearing or deforming in ways that violate physics~\citep{motamed2025generativevideomodelsunderstand, bansal2024videophy, meng2024towards, mei2025confident}. Video models also struggle to follow user instructions~\citep{qin2024worldsimbench, li2025worldmodelbench}, especially in long-duration video generation tasks~\citep{yu2025malt, zhang2025packing, guo2025long}. Further, significant data curation, training, and inference costs and the lack of adequate content safeguards in video generation remain critical challenges~\citep{chen2024safewatchefficientsafetypolicyfollowing, miao2024t2vsafetybenchevaluatingsafetytexttovideo}, hindering broader adoption in robotics. Addressing these challenges constitutes a crucial roadmap for future work, ultimately enabling video models to realize their full potential.

This survey provides an extensive overview of video models, identifying prevailing model architectures, key applications in robotics, and major challenges. We highlight four prominent applications of video models as embodied world models in robotics, including: (i)~robot data generation and action prediction in imitation learning, (ii)~dynamics and rewards modeling in reinforcement learning, (iii)~policy evaluation, and (iv)~visual planning, as illustrated in~\Cref{fig:banner}. Furthermore, we outline important open research challenges impeding trustworthy applications of video models in robotics and present important directions for future research to address these problems.

\p{Comparison to Existing Surveys}
Prior survey papers have covered video generation models broadly, particularly video diffusion models, identifying the prevailing model architectures~\cite{melnik2024video, yu2025survey}, techniques for controllable video generation~\cite{ma2025controllable}, core applications in video generation, video editing, and video understanding~\cite{xing2024survey}. Other surveys~\cite{li2025comprehensive, ding2025understanding, kong20253d, zhang2025generative} provide a broad discussion of world models, presenting a valuable review of research across a wide range of model architectures, such as recurrent neural networks, transformers, language models, and scene reconstruction models, e.g., Gaussian Splatting. 
Prior work also highlights the unified framework for information representation and task formulation provided by video models~\citep{yang2024video} and discuss the progression in the capabilities of video models, through a primarily historical lens, along with potential future capabilities~\citep{yue2025simulatingvisualworldartificial}.
Although relevant for broader understanding, these existing surveys lack a comprehensive review of generative video models as world models in robotics~\cite{li2025comprehensive, ding2025understanding, kong20253d, zhang2025generative} or an extensive discussion of specific applications, associated challenges, and future directions of generative video models in robotics~\cite{melnik2024video, yu2025survey, ma2025controllable, xing2024survey, yue2025simulatingvisualworldartificial}.
In contrast, our survey provides an exhaustive discussion of video generation models in robotics, especially in robot manipulation, which has not been considered by existing surveys unlike areas such as autonomous driving~\cite{zhu2024sora, guan2024world, feng2025survey, li2025drivevla}. 

\p{Research Methodology}
We surveyed the literature for papers related to video generation with particular emphasis on applications to robotics, checking online sources, paper references, and individual suggestions. We considered all publicly accessible papers up to December 2025, aggregating them in a local database. To identify more relevant references, we categorized these papers into foundational methods such as diffusion modeling/flow-matching; application-centric solutions, e.g., in robot policy learning, planning, and policy evaluation; and benchmarks and datasets, including evaluation metrics. Given the broad range of papers on the foundations of video generation, we limit our discussion to seminal papers, along with significant papers introducing new ideas that have been widely adopted by other papers discussed in this survey. In contrast, we include all papers focused on applications of video models to robotics in this survey, especially those where video generation plays a critical role. Likewise, we include all representative video generation benchmark and dataset papers, given their significance to future research. To facilitate further research, we provide a curated bibliography at \url{https://github.com/irom-princeton/awesome-robotics-video-world-model-papers}.

\p{Taxonomy}
We categorize existing literature around its practical applications, framing these domains as the primary vectors of impact for video world models in robotics. This functional taxonomy yields three core contributions. First, it contextualizes state-of-the-art world-modeling methodologies within fundamental robotic paradigms—specifically policy learning, evaluation, and visual planning, thereby rendering the field accessible to a multidisciplinary audience. Second, it systemically identifies the inherent boundaries of current frameworks, illustrating how video-driven approaches can mitigate these limitations. Finally, this taxonomy charts the critical open challenges remaining in the field, detailing the barriers to achieving physics-aligned future prediction, thereby offering a strategic roadmap to guide future research.

\p{Organization}
We discuss the organization of this survey, as summarized in~\Cref{fig:outline}.
In~\Cref{sec:background}, we provide an overview of important concepts that are crucial to understanding generative video modeling. We begin with a discussion of learned world models, highlighting the evolution of these models from (latent) state-based representations to high-fidelity video-based representations. Further, we review diffusion/flow-matching-based modeling highlighting the core principles behind their success. In~\Cref{sec:app_video_models}, we present major applications of video models in robotics, encompassing robot data generation and action prediction in imitation learning, dynamics and rewards modeling in reinforcement learning, policy evaluation, and visual planning. In~\Cref{sec:eval_video_models}, we highlight useful metrics and benchmarks for evaluating video models. Subsequently, we identify open research challenges in~\Cref{sec:challenges_and_future_work} and suggest directions for future research. We conclude the survey in~\Cref{sec:conclusion}.

\input{sections/front_matter/overview}

%% file: sections/front_matter/overview.tex
\begingroup
\hypersetup{linkcolor=black}

\begin{center}
    {
    \begin{adjustbox}{width=0.8\textwidth}
    {
        \begin{tikzpicture}[node distance=2cm]
        
        \node[rotate=90] (level_0) [rect, fill=survey!50] {Video Generation Models in Robotics};
        
        \node (level_1_t0) [rect_wrap, fill=row0_color!90, right of=level_0, xshift=10ex, yshift=46ex] {\hyperref[sec:background]{Background}\onl (\cref{sec:background})};
        \node (level_1_t1) [rect_wrap, fill=row1_color!90, right of=level_0, xshift=10ex, yshift=13ex] {\hyperref[sec:app_video_models]{Applications in Robotics}\onl (\cref{sec:app_video_models})};
        \node (level_1_t2) [rect_wrap, fill=row2_color!90, right of=level_0, xshift=10ex, yshift=-15.5ex] {\hyperref[sec:eval_video_models]{Evaluating Video Models}\onl (\cref{sec:eval_video_models})};
        \node (level_1_t3) [rect_wrap, fill=row3_color!90, right of=level_0, xshift=10ex, yshift=-83ex] {\hyperref[sec:challenges_and_future_work]{Open Challenges and Future Directions}\onl (\cref{sec:challenges_and_future_work})};
        
        \draw [arrow] (level_0.south) -- ++(2.5ex, 0ex) |- (level_1_t0.west);
        \draw [arrow] (level_0.south) -- ++(2.5ex, 0ex) |- (level_1_t1.west);
        \draw [arrow] (level_0.south) -- ++(2.5ex, 0ex) |- (level_1_t2.west);
        \draw [arrow] (level_0.south) -- ++(2.5ex, 0ex) |- (level_1_t3.west);
        
        \node (level_2_t0_0) [rect_large, fill=row0_color!50, right of=level_1_t0, xshift=25ex, yshift=9ex] {\hyperref[sec:bkgd_world_models]{Markovian \mbox{State-Based} World Models}\onl (\cref{sec:bkgd_world_models})};
        \node (level_2_t0_1) [rect_large, fill=row0_color!50, right of=level_1_t0, xshift=25ex, yshift=0ex] {\hyperref[sec:diffusion_video_models]{Diffusion/Flow-Matching Models}\onl (\cref{sec:diffusion_video_models})};
        \node (level_2_t0_2) [rect_large, fill=row0_color!50, right of=level_1_t0, xshift=25ex, yshift=-9ex] {\hyperref[sec:non_diffusion_based_models]{Video JEPA Models}\onl (\cref{sec:non_diffusion_based_models})};
        
        \draw [arrow] (level_1_t0.east) -- ++(2.5ex, 0ex) |- (level_2_t0_0.west);
        \draw [arrow] (level_1_t0.east) -- ++(2.5ex, 0ex) |- (level_2_t0_1.west);
        \draw [arrow] (level_1_t0.east) -- ++(2.5ex, 0ex) |- (level_2_t0_2.west);

        \node (level_2_t1_0) [rect_large, fill=row1_color!50, right of=level_1_t1, xshift=25ex, yshift=13.5ex]  {\hyperref[sec:app_data_gen_il]{Data Generation/\mbox{Backbones} in IL}\onl (\cref{sec:app_data_gen_il})};
        \node (level_2_t1_1) [rect_large, fill=row1_color!50, right of=level_1_t1, xshift=25ex, yshift=4.5ex] {\hyperref[sec:app_rl]{Dynamics/Rewards Models in RL}\onl (\cref{sec:app_rl})};
        \node (level_2_t1_2) [rect_large, fill=row1_color!50, right of=level_1_t1, xshift=25ex, yshift=-4.5ex] {\hyperref[sec:app_visual_plan]{Policy Evaluation}\onl (\cref{sec:app_policy_eval})};
        \node (level_2_t1_3) [rect_large, fill=row1_color!50, right of=level_1_t1, xshift=25ex, yshift=-13.5ex] {\hyperref[sec:app_policy_eval]{Visual Planning}\onl (\cref{sec:app_visual_plan})};

        \draw [arrow] (level_1_t1.east) -- ++(2.5ex, 0ex) |- (level_2_t1_0.west);
        \draw [arrow] (level_1_t1.east) -- ++(2.5ex, 0ex) |- (level_2_t1_1.west);
        \draw [arrow] (level_1_t1.east) -- ++(2.5ex, 0ex) |- (level_2_t1_2.west);
        \draw [arrow] (level_1_t1.east) -- ++(2.5ex, 0ex) |- (level_2_t1_3.west);
        
        \node (level_2_t2_0) [rect_large, fill=row2_color!50, right of=level_1_t2, xshift=25ex, yshift=4.5ex] {\hyperref[subsec:bkgd_metrics]{Metrics}\onl (\cref{subsec:bkgd_metrics})};
        \node (level_2_t2_1) [rect_large, fill=row2_color!50, right of=level_1_t2, xshift=25ex, yshift=-4.5ex] {\hyperref[subsec:bkgd_benchmarks]{Benchmarks}\onl (\cref{subsec:bkgd_benchmarks})};
        \node (level_2_t2_2) [rect_large, fill=row2_color!50, right of=level_1_t2, xshift=25ex, yshift=-13.5ex] {\hyperref[subsec:bkgd_benchmarks]{Axes}\onl (\cref{subsec:eval_axes})};

        \draw [arrow] (level_1_t2.east) -- ++(2.5ex, 0ex) |- (level_2_t2_0.west);
        \draw [arrow] (level_1_t2.east) -- ++(2.5ex, 0ex) |- (level_2_t2_1.west);
        \draw [arrow] (level_1_t2.east) -- ++(2.5ex, 0ex) |- (level_2_t2_2.west);
        
        \node (level_2_t3_0) [rect_large, fill=row3_color!50, right of=level_1_t3, xshift=25ex, yshift=42ex] {\hyperref[subsec:chall_hallucinations_physics_violations]{Hallucinations and \mbox{Violation} of Physics}\onl (\cref{subsec:chall_hallucinations_physics_violations})};
        \node (level_2_t3_1) [rect_large, fill=row3_color!50, right of=level_1_t3, xshift=25ex, yshift=31.5ex]{\hyperref[subsec:chall_uq]{Uncertainty Quantification}\onl (\cref{subsec:chall_uq})};
        \node (level_2_t3_2) [rect_large, fill=row3_color!50, right of=level_1_t3, xshift=25ex, yshift=22.5ex] {\hyperref[subsec:chall_instr_follow]{Instruction Following}\onl (\cref{subsec:chall_instr_follow})};
        \node (level_2_t3_3) [rect_large, fill=row3_color!50, right of=level_1_t3, xshift=25ex, yshift=13.5ex] {\hyperref[subsec:chall_eval_video_models]{Evaluating Video Models}\onl (\cref{subsec:chall_eval_video_models})};
        \node (level_2_t3_4) [rect_large, fill=row3_color!50, right of=level_1_t3, xshift=25ex, yshift=4.5ex] {\hyperref[subsec:chall_safe_content]{Safe Content Generation}\onl (\cref{subsec:chall_safe_content})};
        \node (level_2_t3_5) [rect_large, fill=row3_color!50, right of=level_1_t3, xshift=25ex, yshift=-4.5ex] {\hyperref[subsec:chall_safe_robot_interactions]{Safe Robot Interactions}\onl (\cref{subsec:chall_safe_robot_interactions})};
        \node (level_2_t3_6) [rect_large, fill=row3_color!50, right of=level_1_t3, xshift=25ex, yshift=-13.5ex] {\hyperref[subsec:chall_action_estimation]{Action Estimation}\onl (\cref{subsec:chall_action_estimation})};
        \node (level_2_t3_7) [rect_large, fill=row3_color!50, right of=level_1_t3, xshift=25ex, yshift=-22.5ex] {\hyperref[subsec:long_video_generation]{Long-Horizon Video Generation}\onl (\cref{subsec:long_video_generation})};
        \node (level_2_t3_8) [rect_large, fill=row3_color!50, right of=level_1_t3, xshift=25ex, yshift=-31.5ex] {\hyperref[subsec:chall_data_curation]{Data Curation Costs}\onl (\cref{subsec:chall_data_curation})};
        \node (level_2_t3_9) [rect_large, fill=row3_color!50, right of=level_1_t3, xshift=25ex, yshift=-40.5ex] {\hyperref[subsec:chall_training_overhead]{Training and Inference Costs}\onl (\cref{subsec:chall_training_overhead})};

        \draw [arrow] (level_1_t3.east) -- ++(2.5ex, 0ex) |- (level_2_t3_0.west);
        \draw [arrow] (level_1_t3.east) -- ++(2.5ex, 0ex) |- (level_2_t3_1.west);
        \draw [arrow] (level_1_t3.east) -- ++(2.5ex, 0ex) |- (level_2_t3_2.west);
        \draw [arrow] (level_1_t3.east) -- ++(2.5ex, 0ex) |- (level_2_t3_3.west);
        \draw [arrow] (level_1_t3.east) -- ++(2.5ex, 0ex) |- (level_2_t3_4.west);
        \draw [arrow] (level_1_t3.east) -- ++(2.5ex, 0ex) |- (level_2_t3_5.west);
        \draw [arrow] (level_1_t3.east) -- ++(2.5ex, 0ex) |- (level_2_t3_6.west);
        \draw [arrow] (level_1_t3.east) -- ++(2.5ex, 0ex) |- (level_2_t3_7.west);
        \draw [arrow] (level_1_t3.east) -- ++(2.5ex, 0ex) |- (level_2_t3_8.west);
        \draw [arrow] (level_1_t3.east) -- ++(2.5ex, 0ex) |- (level_2_t3_9.west);
        
        \node (level_3_t0_0) [rect_medium, fill=row0_color!20, align=left, right of=level_2_t0_0, xshift=27ex, yshift=0ex] { \hspace{0.2ex} [\citealp{hafner2023mastering}, \citealp{hansen2023td}, \citealp{zhou2024dino}, \citealp{vondrick2016generating}, \citealp{babaeizadeh2017stochastic}, \citealp{wu2024ivideogpt}, \citealp{nakamura2025generalizing}, \citealp{yin2025womap}]}; %
        \node (level_3_t0_1) [rect_medium, fill=row0_color!20, align=left, right of=level_2_t0_1, xshift=27ex, yshift=0ex] {\hspace{0.2ex}  [\citealp{ho2020denoising}, \citealp{lipman2023flowmatchinggenerativemodeling}, \citealp{brooks2024video}, \citealp{deepmind2025genie3}, \citealp{deepmind_veo3_techreport}, \citealp{wan2025wan}, \citealp{agarwal2025cosmos}, \citealp{ho2022video}, \citealp{blattmann2023stable}]};
        \node (level_3_t0_2) [rect_medium, fill=row0_color!20, align=left, right of=level_2_t0_2, xshift=27ex, yshift=0ex] {\hspace{0.2ex}  [\citealp{assran2023self}, \citealp{bardes2024revisiting}, \citealp{garrido2025intuitivephysicsunderstandingemerges}, \citealp{assran2025v}]};
        
        \draw [arrow] (level_2_t0_0.east) -- (level_3_t0_0.west);
        \draw [arrow] (level_2_t0_1.east) -- (level_3_t0_1.west);
        \draw [arrow] (level_2_t0_2.east) -- (level_3_t0_2.west);

        \node (level_3_t1_0) [rect_medium, fill=row1_color!20, align=left, right of=level_2_t1_0, xshift=27ex, yshift=0ex] {\hspace{0.2ex}  [\citealp{jang2025dreamgen}, \citealp{patel2025robotic}, \citealp{qiu2025lucibot}, \citealp{bharadhwaj2024gen2act}, \citealp{soni2024videoagent}, \citealp{liang2024dreamitate}, \citealp{li2025unified}, \citealp{zhang2025dreamvla}]};
        \node (level_3_t1_1) [rect_medium, fill=row1_color!20, align=left, right of=level_2_t1_1, xshift=27ex, yshift=0ex] {\hspace{0.2ex}  [\citealp{hafner2025training}, \citealp{xiao2025world},  \citealp{escontrela2023video}, \citealp{huang2024diffusion}, \citealp{jiang2025enerverse}, \citealp{luo2024grounding}]};
        \node (level_3_t1_2) [rect_medium, fill=row1_color!20, align=left, right of=level_2_t1_2, xshift=27ex, yshift=0ex] {\hspace{0.2ex}  [\citealp{quevedo2025worldgymworldmodelenvironment}, \citealp{guo2025ctrl}, \citealp{1x_world}, \citealp{veorobotics2025}, \citealp{li2025worldeval}, \citealp{zhu2024irasim}, \citealp{tseng2025scalable}]};
        \node (level_3_t1_3) [rect_medium, fill=row1_color!20, align=left, right of=level_2_t1_3, xshift=27ex, yshift=0ex] {\hspace{0.2ex}  [\citealp{li2025novaflow}, \citealp{du2023videolanguageplanning}, \citealp{gao2024flip}, \citealp{ebert2018visual}, \citealp{bu2024closed}, \citealp{yang2025mindjourney}, \citealp{cen2024using}, \citealp{black2023zero}]};
        
        \draw [arrow] (level_2_t1_0.east) -- (level_3_t1_0.west);
        \draw [arrow] (level_2_t1_1.east) -- (level_3_t1_1.west);
        \draw [arrow] (level_2_t1_2.east) -- (level_3_t1_2.west);
        \draw [arrow] (level_2_t1_3.east) -- (level_3_t1_3.west);
        
        \node (level_3_t2_0) [rect_medium, fill=row2_color!20, align=left, right of=level_2_t2_0, xshift=27ex, yshift=0ex] {\hspace{0.2ex}  [\citealp{Hore2010PSNR}, \citealp{wang2004image}, \citealp{zhang2018unreasonableeffectivenessdeepfeatures}, \citealp{radford2021learning}, \citealp{unterthiner2018towards}, \citealp{gibson1951perception}]};
        \node (level_3_t2_1) [rect_medium, fill=row2_color!20, align=left, right of=level_2_t2_1, xshift=27ex, yshift=0ex] {\hspace{0.2ex} [\citealp{bansal2024videophy}, \citealp{meng2024towards}, \citealp{liu2024evalcrafter}, \citealp{huang2024vbench}, \citealp{sun2025t2v},  \citealp{tian2023control}]};
        \node (level_3_t2_2) [rect_medium, fill=row2_color!20, align=left, right of=level_2_t2_2, xshift=27ex, yshift=0ex] {\hspace{0.2ex} [\citealp{tian2023control}, \citealp{guo2025ctrl}, \citealp{wang2025language}]};

        \draw [arrow] (level_2_t2_0.east) -- (level_3_t2_0.west);
        \draw [arrow] (level_2_t2_1.east) -- (level_3_t2_1.west);
        \draw [arrow] (level_2_t2_2.east) -- (level_3_t2_2.west);
        
        \node (level_3_t3_0) [rect_medium, fill=row2_color!20, align=left, right of=level_2_t3_0, xshift=27ex, yshift=0ex] {\hspace{0.2ex}  [\citealp{motamed2025generativevideomodelsunderstand}, \citealp{li2025worldmodelbench}, \citealp{lin2025exploring}, \citealp{kang2024far}, \citealp{zhang2025thinkdiffusellmsguidedphysicsaware}]};
        \node (level_3_t3_1) [rect_medium, fill=row2_color!20, align=left, right of=level_2_t3_1, xshift=27ex, yshift=0ex] {\hspace{0.2ex} [\citealp{mei2025confident}, \citealp{mei2025world}]};
        \node (level_3_t3_2) [rect_medium, fill=row2_color!20, align=left, right of=level_2_t3_2, xshift=27ex, yshift=0ex] {\hspace{0.2ex} [\citealp{fang2024vimi}, \citealp{xing2025aid}, \citealp{li2025trainingfreeguidancetexttovideogeneration}, \citealp{zhang2024interactivevideousercentriccontrollablevideo}, \citealp{wang2025atitrajectoryinstructioncontrollable}, \citealp{yuan2024instructvideo}]};
        \node (level_3_t3_3) [rect_medium, fill=row2_color!20, align=left, right of=level_2_t3_3, xshift=27ex, yshift=0ex] {\hspace{0.2ex} [\citealp{qin2024worldsimbench}, \citealp{li2025worldmodelbench}, \citealp{liu2024evalcrafter}]};
        \node (level_3_t3_4) [rect_medium, fill=row2_color!20, align=left, right of=level_2_t3_4, xshift=27ex, yshift=0ex] {\hspace{0.2ex} [\citealp{chen2024safewatchefficientsafetypolicyfollowing}, \citealp{miao2024t2vsafetybenchevaluatingsafetytexttovideo}, \citealp{yoon2025safreetrainingfreeadaptiveguard}]};
        \node (level_3_t3_5) [rect_medium, fill=row2_color!20, align=left, right of=level_2_t3_5, xshift=27ex, yshift=0ex] {\hspace{0.2ex} [\citealp{veorobotics2025}, \citealp{nakamura2025generalizing}, \citealp{seo2025uncertaintyawarelatentsafetyfilters}]};
        \node (level_3_t3_6) [rect_medium, fill=row2_color!20, align=left, right of=level_2_t3_6, xshift=27ex, yshift=0ex] {\hspace{0.2ex} [\citealp{deepmind2025genie3}, \citealp{ye2024latent}, \citealp{chen2025villa}, \citealp{pathak2018zero}, \citealp{tan2025anypos}]};
        \node (level_3_t3_7) [rect_medium, fill=row2_color!20, align=left, right of=level_2_t3_7, xshift=27ex, yshift=0ex] {\hspace{0.2ex} [\citealp{yu2025malt}, \citealp{zhang2025packing}, \citealp{guo2025long}, \citealp{NEURIPS2024_2aee1c41}, \citealp{dalal2025one}, \citealp{zhang2025test}]};
        \node (level_3_t3_8) [rect_medium, fill=row2_color!20, align=left, right of=level_2_t3_8, xshift=27ex, yshift=0ex] {\hspace{0.2ex} [\citealp{agarwal2025cosmos}, \citealp{yang2024cogvideox}, \citealp{chen2024panda}, \citealp{tan2024vidgen}, \citealp{nan2024openvid}, \citealp{wang2025koala}]};
        \node (level_3_t3_9) [rect_medium, fill=row2_color!20, align=left, right of=level_2_t3_9, xshift=27ex, yshift=0ex] {\hspace{0.2ex} [\citealp{hafner2025training}, \citealp{ko2025implicitstateestimationvideo}, \citealp{blattmann2023align}, \citealp{peng2025open}, \citealp{frans2024one}, \citealp{song2023consistency}]};
        
        \draw [arrow] (level_2_t3_0.east) -- (level_3_t3_0.west);
        \draw [arrow] (level_2_t3_1.east) -- (level_3_t3_1.west);
        \draw [arrow] (level_2_t3_2.east) -- (level_3_t3_2.west);
        \draw [arrow] (level_2_t3_3.east) -- (level_3_t3_3.west);
        \draw [arrow] (level_2_t3_4.east) -- (level_3_t3_4.west);
        \draw [arrow] (level_2_t3_5.east) -- (level_3_t3_5.west);
        \draw [arrow] (level_2_t3_6.east) -- (level_3_t3_6.west);
        \draw [arrow] (level_2_t3_7.east) -- (level_3_t3_7.west);
        \draw [arrow] (level_2_t3_8.east) -- (level_3_t3_8.west);
        \draw [arrow] (level_2_t3_9.east) -- (level_3_t3_9.west);

        \end{tikzpicture}
        }
    \end{adjustbox}
    }
    
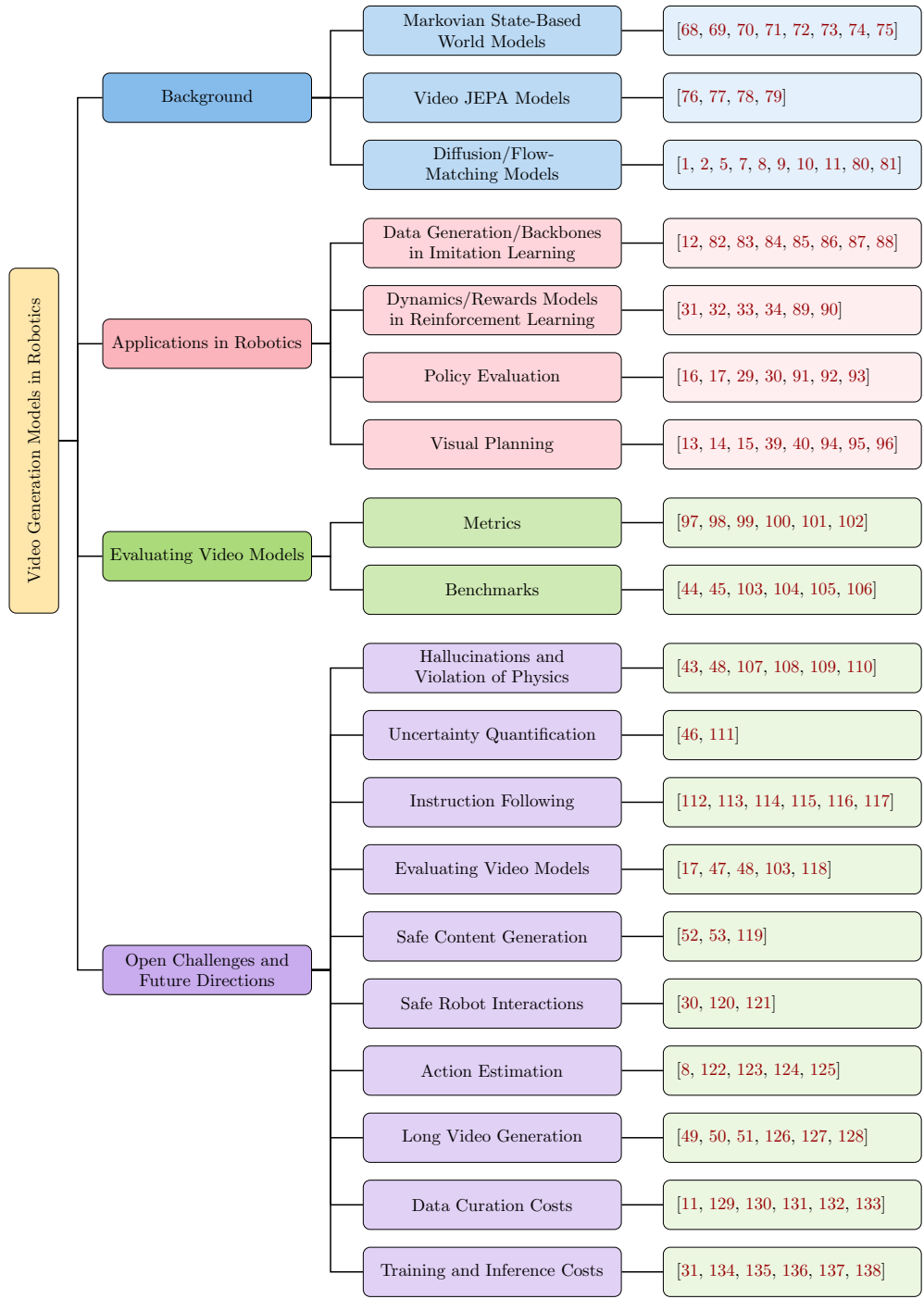
\captionof{figure}{\textbf{Organization.} The organization of this survey, including background material, taxonomy of robotics applications, evaluation metrics and benchmarks, and open challenges and directions for future research.}
    \label{fig:outline}
\end{center}
\endgroup

%% file: sections/background/background.tex
\section{Background}
\label{sec:background}

In this section, we provide relevant background concepts that are essential to understanding the material presented in this survey. We begin with a brief introduction of world models, followed by an overview of video models. 
Specifically, we categorize video models broadly into non-diffusion-based video models and diffusion/flow-matching-based video models. State-of-the-art video models generally utilize diffusion or flow-matching for high-fidelity video synthesis. 
As a result, we provide a more comprehensive discussion of these models.

\input{sections/background/world_models/bkgd_world_models_nondiffusion}
\input{sections/background/diffusion_models/diffusion_video_models}

\input{sections/background/non_diffusion_models/other_latent_models}

%% file: sections/background/world_models/bkgd_world_models_nondiffusion.tex
\subsection{Learned World Models}
\label{sec:bkgd_world_models}

Many robotics algorithms require a model of the robot's environment to efficiently learn policies that are effective in the real-world, especially when real-world interactions are prohibitively costly or unsafe. For example, reinforcement learning (RL) methods~\cite{rajeswaran2017learning, matas2018sim} generally require a large number of interactions between an agent and its environment to learn useful behaviors, which is typically expensive in the real-world. World models enable scalable data collection for training these policies with little to no real-world interaction.
At their core, world models predict the evolution of the environment of an agent due to interactions. 
Traditionally, physics-based simulators~\cite{todorov2012mujoco, coumans2016pybullet} have been utilized as world models for predicting the dynamical effects of a robot's actions.
However, physics-based simulators typically utilize simplified physics engines that approximate physical laws for computational feasibility, introducing inductive biases that limit their realism, e.g., in simulating non-rigid objects. Importantly, these approximations often contribute to the sim-to-real gap in robotics, impeding successful transfer of simulation-trained robot policies to the real-world. Moreover, robot manipulation tasks have become increasingly complex, exacerbating these challenges.

Learned world models~\cite{ha2018recurrent, ha2018world} have emerged as an effective solution to these challenges.
We can broadly classify learned world models into two categories: \emph{Markovian state-based} world models and \emph{video} world models.

\p{Markovian State-Based World Models}
A state refers to a description of the environment of an agent at a given time step, which could include its RGB observations, proprioception, or latent embeddings of these observations at past and current time steps. Markovian state-based world models assume that the future evolution of the agent's environment only depends on the agent's state $s_{t}$ at the current time step $t$ and its action $a_{t}$, which represents either a physical action or a latent action.
Markovian state-based world models are trained to predict the future state $s_{t+1}$, given the current state $s_t$ and action $a_t$: ${s_{t+1} \sim p_\eta(s_{t+1} | s_t, a_t),}$
where $p_\eta$ represents the dynamics predictor with parameters $\eta$. 
These world models generally operate in latent space and are composed of an encoder for embedding observations into a latent space, dynamics predictor, and rewards predictor, given by:
\begin{equation}
    \text{Encoder: } \ s_t\sim\mathcal E_\gamma(s_t | o_t ), \quad
    \text{Dynamics: } \ \hat s_{t+1}\sim p_\eta(\hat s_{t+1} | s_t, a_t), \quad
    \text{Rewards: } \ \hat r_{t+1}\sim p_\zeta (\hat r_{t+1} | \hat s_{t+1}),
\end{equation}
with parameters $\gamma$ and $\zeta$ for the encoder and rewards predictor, respectively.

Prior work~\cite{hafner2020dreamcontrollearningbehaviors, hafner2020mastering,hafner2023mastering, hansen2023td, mendonca2023structured, wu2023pre} has largely parameterized the dynamics predictor $p_\eta$ using recurrent neural networks (RNNs) or recurrent state-space models (RSSMs).
However, more recent works employ transformers~\cite{zhou2024dino} and diffusion in pixel space~\cite{alonso2024diffusion, yang2023learning, ding2024diffusion} or latent space~\cite{huang2025ladi, bar2025navigation} for more expressive dynamics prediction.
In general, world models~\citep{nakamura2025generalizing, yin2025womap, hafner2020dreamcontrollearningbehaviors, wu2023daydreamer, zhou2024dino, chandra2025diwa, zhang2025dreamvla} are trained to minimize the error between the predicted and ground-truth next state, either in latent space or reconstructed pixel space using the mean-squared error (MSE) loss function~\cite{zhou2024dino, alonso2024diffusion} or the Kullback-Leibler (KL) divergence loss~\cite{hafner2023mastering}.
We refer interested readers to~\citep{li2025comprehensive} for a detailed review of these models.

\p{Video World Models}
To model the world, video models learn spatiotemporal mappings that capture the evolution of an environment over space and time, without explicitly modeling a Markovian state, in contrast to Markovian state-based world models.
In general, video models apply non-linear transformations to the patches or pixels of a video frame to propagate the environment dynamics.
Early research in video prediction applied spatial transformations to pixels to generate future frames \cite{finn2016unsupervised, jia2016dynamic}. However, these transformations capture only local perturbations of the frames, limiting their expressiveness in more complex video generation, e.g., with dynamic backgrounds. 
Other methods \cite{vondrick2016generating, clark2019adversarial} extend generative adversarial networks (GANs) \cite{goodfellow2014generative} to video prediction for higher video fidelity. However, the susceptibility of GANs to mode collapse limits their ability to model diverse potential video evolution paths.

Subsequent work \cite{babaeizadeh2017stochastic, lee2018stochastic, walker2021predicting, babaeizadeh2021fitvid, wu2024ivideogpt} uses variational inference via variational autoencoders \cite{kingma2013auto} to learn a latent distribution over pixels, explicitly encoding the stochasticity in video prediction within the latent distribution. 
Given the success of transformers in language modeling, the work in \cite{weissenborn2019scaling} introduces the video transformer, achieving higher-quality video generation through autoregressive predictions.
However, powerful autoregressive decoders introduce the challenge of latent collapse, where the learned model fails to efficiently use the latent space. Vector-quantized variational autoencoders (VQ-VAEs)~\cite{van2017neural, wan2025wan, agarwal2025cosmos, walker2021predicting, wu2024ivideogpt, yan2021videogptvideogenerationusing, bruce2024genie} address this issue by quantizing the latent representations into discrete codes in a codebook.
VQ-VAEs are typically pre-trained on large-scale image and video datasets with a reconstruction objective on compact per-frame latent embeddings achieved through spatial and temporal compression. 

Despite the resulting performance gains, these video models lack the ability to capture complex interactions within an environment, limiting their applications. 
These limitations have been addressed by diffusion and flow-based matching, enabling video models to learn highly expressive representations of the physical world.
We provide a detailed discussion of these video models in the subsequent subsections, along with a brief overview of joint-embedding predictive architectures, which seek to learn world models from video data without predicting pixel-level visual details.

%% file: sections/background/diffusion_models/diffusion_video_models.tex
\subsection{Diffusion/Flow-Matching Video Models}
\label{sec:diffusion_video_models}

\begin{figure}
    \centering
    \includegraphics[width=0.8\linewidth]{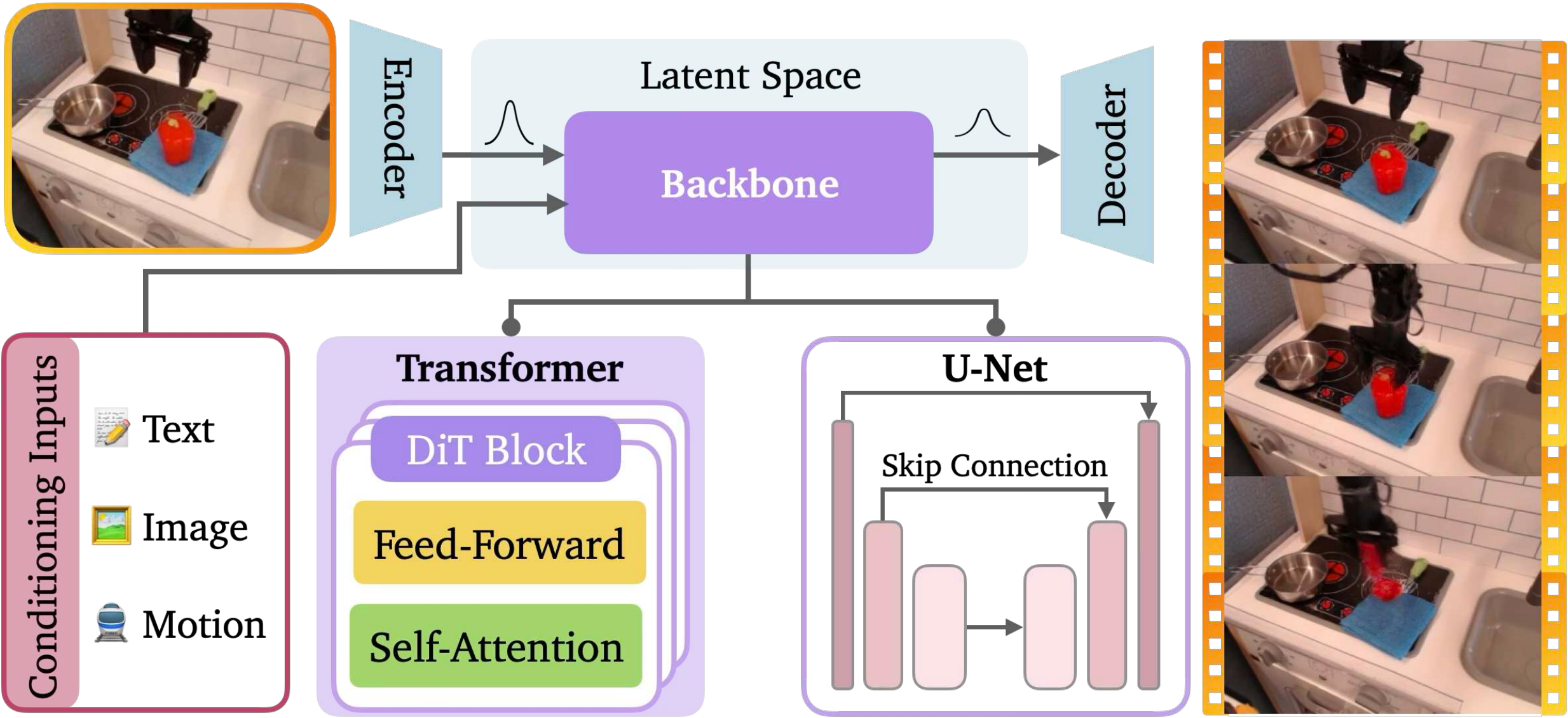}
    \caption{\textbf{Diffusion Video Model Architectures.} Diffusion/Flow-matching has emerged as the dominant model architecture for training photorealistic controllable video models that can be steered using text, image, and other conditioning inputs. These models broadly utilize diffusion transformers (DiTs) or U-Nets to learn important interpendencies across space and time within a compact latent space.}
    \label{fig:architecture}
\end{figure}

The emergence of diffusion modeling~\citep{ho2020denoising} and flow-matching~\citep{lipman2023flowmatchinggenerativemodeling} has transformed the SOTA in image and video generation~\citep{dhariwal2021diffusion, kong2020diffwave}, achieving high-fidelity video generation that capture fine-grained realistic and cinematic effects. 
Given the similarity between diffusion modeling and flow-matching, we limit our discussion to diffusion models for simplicity.
We refer readers to~\citep{lipman2023flowmatchinggenerativemodeling} for a more detailed presentation of flow-matching.
Diffusion models \cite{ho2020denoising, song2022denoisingdiffusionimplicitmodels} have gained prominence as a powerful formulation for generative models that synthesize data by modeling an iterative denoising process. 
Given data $\mathbf{x}_0 \sim q(\mathbf{x}_0)$, the \emph{forward process} (also called the diffusion process) progressively corrupts clean data $\mathbf{x}_0$ with Gaussian noise over $T$ time steps under some noise schedule $\{\beta_t\}_{t=1}^T$, given by: ${q(\mathbf{x}_t | \mathbf{x}_{t-1}) = \mathcal{N}\left(\mathbf{x}_t;\sqrt{1 - \beta_t}\, \mathbf{x}_{t-1}, \, \beta_t \mathbf{I}\right).}$
Conversely, the \emph{reverse process} iteratively %
removes the noise added to the images in the forward process.
These transitions are supervised via a noise-based or velocity-based loss function~\citep{ho2020denoising}, given by:
${L_{\boldsymbol{\epsilon}} = \mathbb{E}_{\mathbf{x}_0, t, \boldsymbol{\epsilon}} 
    \left[
    \left\|
    \boldsymbol{\epsilon} - \boldsymbol{\epsilon}_\theta(\mathbf{x}_t, t)
    \right\|^2
    \right]}$
and
${L_{\boldsymbol{v}} =
    \mathbb{E}_{\mathbf{x}_0, t, \boldsymbol{v}_t}
    \left[
    \left\|
    \boldsymbol{v}_t - \boldsymbol{v}_\theta(\mathbf{x}_t, t)
    \right\|^2
    \right]}$,
respectively.

Early diffusion models~\citep{ho2022video} operate directly in the pixel space, which is prohibitively expensive at high resolutions. To overcome this limitation, recent approaches adopt a latent diffusion paradigm, where observations are first compressed into a lower-dimensional latent space using a variational autoencoder (VAE)~\citep{rombach2022high, blattmann2023stable, kong2024hunyuanvideo, wan2025wan, agarwal2025cosmos} or pretrained encoders~\citep{baldassarre2025back} prior to the diffusion process.
Diffusion models are broadly parameterized using U-Nets~\citep{ronneberger2015u, cciccek20163d} or transformers~\citep{peebles2023scalable, chen2024gentron}.
Early works on video diffusion~\cite{ho2022video, wu2023tune, blattmann2023align, blattmann2023stable} extend 2D U-Nets to the spatiotemporal domain by lifting 2D convolutions to 3D or by introducing temporal attention modules to jointly model spatial and temporal dependencies.
More recent work~\citep{peebles2023scalable, chen2024gentron} replace the traditional U-Net backbone with a diffusion transformers (DiT), leveraging the expressivity of self-attention mechanisms. 
This design trades the U-Net’s inductive biases toward locality and translation equivariance for a more flexible capacity to model long-range dependencies and higher-order semantic relationships, which are important in temporally-coherent long-video generation, an essential requirement in robotics.
Consequently, SOTA video models~\citep{chen2024gentron, wan2025wan, RunwayGen3, brooks2024video,
deepmind_veo3_techreport} utilize DiT-based architectures.

Although inherently unconditional, diffusion models must be conditioned on tasks or robot actions for applications to robotics, which is achieved through classifier-based and classifier-free guidance. Classifier-based guidance~\cite{dhariwal2021diffusion} trains a classifier that learns to predict an attribute (label) $y$ from a noisy sample $x_t$, denoted as $p_\phi(y \mid x_t)$, which provides a steering mechanism at inference time. To circumvent the high computation costs and training instability associated with this approach, classifier-free guidance~\citep{ho2022classifier, xing2024dynamicrafter, yin2023dragnuwafinegrainedcontrolvideo, wang2024motionctrl, guo2024animatediffanimatepersonalizedtexttoimage} trains a joint conditional and unconditional model which is steered towards desirable outputs via a modulated conditioning signal.

Conditioning is typically achieved through channel concatenation, cross-attention, or adaptive normalization. While channel concatenation enforces spatially-aligned interactions by augmenting the inputs or latent representations along the channel dimension, cross-attention captures primarily semantic, non-spatial interactions between intermediate video features (queries) and the conditioning signal (keys and values)~\citep{liu2024understandingcrossselfattentionstable}, e.g., with text inputs. In contrast, adaptive normalization methods such as AdaLN or FiLM \cite{perez2017filmvisualreasoninggeneral, xu2019understandingimprovinglayernormalization} provide global attribute control (e.g., target frame rate or motion intensity) through scale and shift modulation of normalization layers.
These mechanisms are directly employed in text-to-video (T2V) generation~\cite{ho2022imagen, singer2022make} via cross-attention, image-to-video (I2V) generation~\citep{blattmann2023stable, xing2024dynamicrafter} with frame-level input concatenation and text-video cross-attention, and action-conditioned (AC) motion/trajectory-guided generation~\citep{gillman2025forcepromptingvideogeneration, zhang2023addingconditionalcontroltexttoimage, agarwal2025cosmos, guo2025ctrl, quevedo2025worldgymworldmodelenvironment, kim2025freeaction, veorobotics2025} using coordinate maps, optical flow fields, keypoint heatmaps, applied forces, or robot joint/end-effector positions.

%% file: sections/background/non_diffusion_models/other_latent_models.tex
\subsection{Video Joint-Embedding Predictive Architecture Models}
\label{sec:non_diffusion_based_models}
Here, we discuss methods based on joint-embedding predictive architectures (JEPAs)~\citep{assran2023self, bardes2024revisiting, assran2025v} that learn world models from internet-scale videos for future prediction, video understanding, and planning in latent space.
In contrast to diffusion/flow-matching video models, which prioritize photorealistic video generation, video JEPA models~\citep{bardes2024revisiting, assran2025v} seek to learn effective latent representations through self-supervised training on video data. 
JEPAs build on contrastive learning approaches that identify salient characteristic features of (image) inputs to distinguish between similar and dissimilar (image) pairs, where similar examples are typically generated via data augmentation, e.g., through cropping, resizing, color distortions, or noise injection. These augmented inputs are mapped to latent embeddings using a learned model, which is trained using the InfoNCE loss function: ${\mathcal{L}_{i,j} = -\log\frac{\exp(\text{sim}(z_i, z_j)/\tau)}{\sum_{k=1}^{2N}\mathbbm{1}_{k\neq i}\exp(\text{sim}(z_i, z_k)/\tau)},}$
where $\text{sim}(\cdot)$ corresponds to the normalized dot product of the two vectors, given a positive image pair ${(x_i, x_j)}$ from a dataset of $N$ image pairs.

To minimize the risk of representation collapse often associated with contrastive learning, JEPA utilizes an asymmetric architecture for embedding input images, comprising a student encoder, a target encoder, and a predictor.
Video joint-embedding predictive architecture (V-JEPA) models \cite{bardes2024revisiting, assran2025v} build upon this foundation for video generation, predicting spatially and temporally masked features within the latent space. By making predictions within the latent space as opposed to pixel space, the model is encouraged to predict higher-level representations rather than pixel-level information \cite{garrido2025intuitivephysicsunderstandingemerges}. This abstract understanding allows the model to build an internal world model, enabling it to also make predictions about future states and plan without task-specific training or reward, which is often unavailable in robotics.
For conditional video generation, the model is trained with the loss function: ${\min_{\theta, \phi, \Delta_y} ||P_{\phi}(\Delta_y, E_\theta(x)) - \text{sg}(E_{\bar\theta}(y))||_1,}$, with future latent embeddings from the predictor network $P_{\phi}$ and frame embeddings from the encoder $E_{\theta}$ and exponentially-moving variant $E_{\bar\theta}$ at the current timestep,
where $\text{sg}(\cdot)$ denotes the stop-gradient operator.
After $E_\theta$ has been trained, it can be frozen and used to train a transition model with additional action-labeled trajectory data \cite{assran2025v}. Although JEPA models are more robust to collapse, they can still be prone to representation collapse~\citep{xing2025critiquesworldmodels}, although they are more robust compared to prior approaches. Regularizers \cite{balestriero2025lejepaprovablescalableselfsupervised, bardes2022vicregvarianceinvariancecovarianceregularizationselfsupervised} have been shown to help prevent posterior collapse, but are difficult to tune and interpret~\citep{xing2025critiquesworldmodels}.

%% file: sections/applications/applications.tex
\section{Applications of Video Models in Robotics}
\label{sec:app_video_models}
In this section, we explore applications of video models as high-fidelity world models in robotics.
\input{sections/applications/app_world_modeling}

\input{sections/applications/app_imitation_learning}

\input{sections/applications/app_rl}

\input{sections/applications/app_policy_eval}

\input{sections/applications/app_planning}

%% file: sections/applications/app_world_modeling.tex
World models capture the evolution of an environment under the effects of actions applied by an agent. However, modeling physical interactions is extremely challenging, given the stochastic nature of dynamical interactions, which is often not fully described by simplified physical laws. 
Although early world models~\cite{hansen2023td, hafner2020dreamcontrollearningbehaviors} effectively learn high-level scene dynamics, they struggle with modeling intricate dynamical changes, especially with high-fidelity. Notably, in fine-grained interaction tasks in robotics, such as dexterous manipulation, these models often fail to predict the effect of subtle changes in robot actions that have significant impacts on the success of such tasks. 
For high-fidelity world modeling, these models generally require lots of training data centered around such dynamical events for effective supervision, which is often difficult to collect in many practical applications.
More traditional physics-based simulators function as world models; however, their capabilities are often limited by restrictive assumptions built on simplified object models and dynamics models. For example, physics-based simulators typically struggle with deformable-body simulations, which require higher-fidelity dynamics model. Importantly, these assumptions contribute to the sim-to-real gap faced by physics-based simulators, e.g., through the use of primitive object shapes, appearance and visual conditions.
Recent work in robotics seeks to address these limitations by leveraging video models as \emph{embodied world models}.

\begin{figure}[t]
    \centering
    \includegraphics[width=0.9\linewidth]{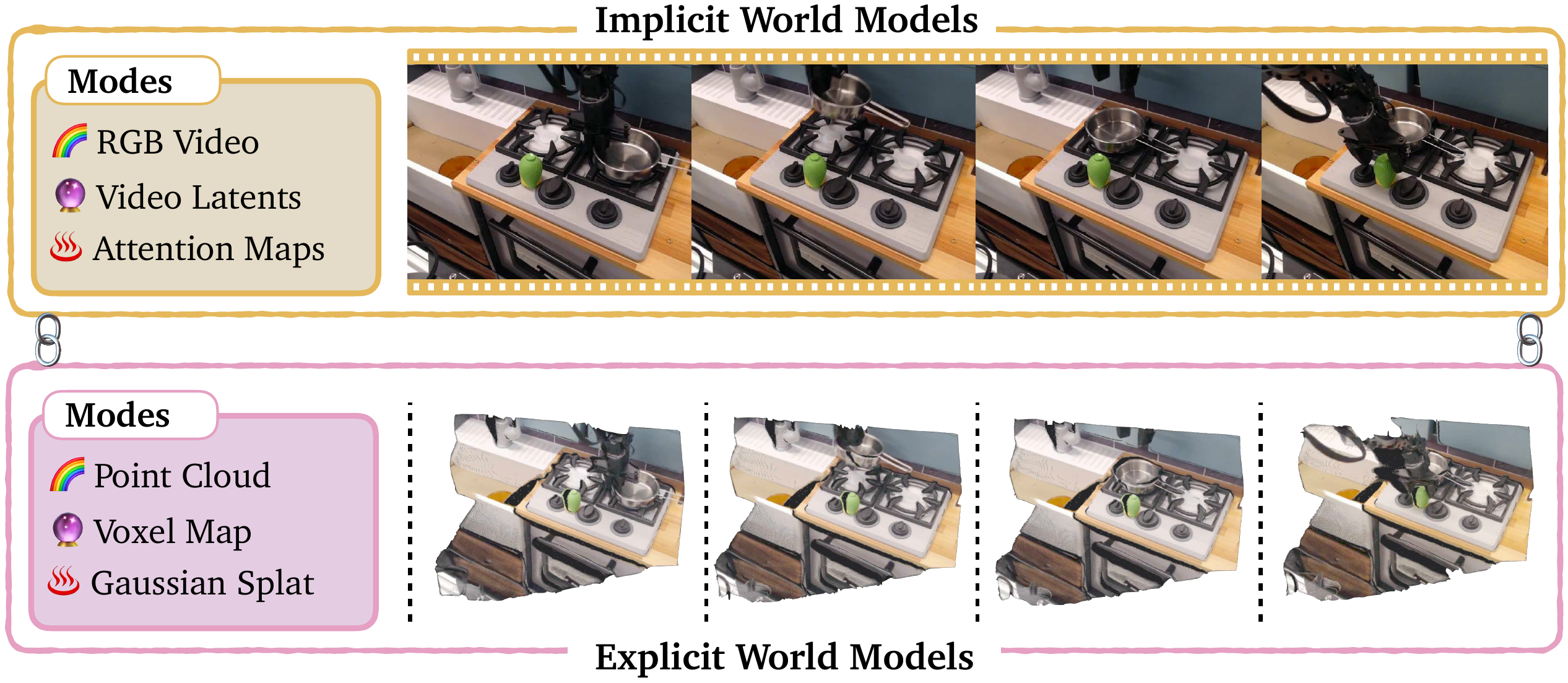}
    \caption{\textbf{Video Models for Embodied World Modeling.} Video models provide high-quality representations of the physical world, which could be implicit (e.g., latent and video representations) or explicit (e.g., point clouds and Gaussian Splatting models).}
    \label{fig:embodied_wm}
\end{figure}

Video models offer a promising path to high-fidelity physical world simulation, without a prohibitive demand for large-scale action-labeled datasets. Recent approaches \cite{xiang2024pandora, huang2025enerverse, team2025aether, huang2025vid2world, deepmind2025genie3} fine-tune pre-trained video models to autoregressively synthesize 4D scene representations, generating new future frames conditioned on past generated ones. Based on their underlying scene representations, video world models can be classified into two categories, based on the maintained scene representation, namely: \emph{implicit} and \emph{explicit} video world models, as illustrated in~\Cref{fig:embodied_wm}.

\p{Implicit video world models}
In implicit video world models, the 3D scene representation is encoded exclusively within the video model, without any external representation.
As such, visualization of the evolving scene can only be achieved by generating videos from the video model.
Implicit representations circumvent the rigid inductive biases common in alternative approaches, yielding superior generalization with fewer assumptions while eliminating the need for auxiliary sensor-based supervision, such as depth maps. Recent works extensively leverage this flexibility for more versatile video generation across both scene-centric~\citep{xiang2024pandora, kim2025freeaction, qian2025wristworld} and object-centric~\cite{villar2025playslot} paradigms, often incorporating fine-grained language steering~\cite{huang2025vid2world} via causal attention and diffusion forcing~\cite{NEURIPS2024_2aee1c41}.
Within these frameworks, conditioning modalities diverge: whereas some approaches \cite{xiang2024pandora} rely solely on text prompts, others \cite{huang2025enerverse, qian2025wristworld, huang2025vid2world, finn2016unsupervised, luo2025solving, zhao2025taste} fine-tune pre-trained models on control-centric robot datasets (e.g., Bridge \cite{ebert2021bridge}, DROID \cite{khazatsky2024droid}) to achieve action-conditioned synthesis.
Nonetheless, lacking structural constraints, these implicit methods demand substantially more training data to internalize underlying physical laws.

\p{Explicit video world models}
In contrast to implicit approaches, explicit video world models ground future predictions in a concrete 3D/4D spatial representation derived from multi-view frames or geometric depth maps, modalities that are often unavailable in real-world settings. To circumvent this data scarcity, recent explicit frameworks~\citep{huang2025enerverse, liao2025genie} leverage cost-effective alternatives, such as simulation data, to scale training on multi-view video data conditioned on text and image inputs. However, synthesized multi-view frames typically yield a sparse representation of the physical world that may prove insufficient for fine-grained robotic manipulation. For finer spatial granularity, contemporary methods integrate structural graphics primitives, utilizing dense 3D Gaussian Splatting fields~\cite{kerbl20233d}, camera ray map back-projection~\citep{team2025aether}, or volumetric voxel grids~\cite{zhou2025learning}.

Given their growing capabilities, video models enable a broad range of downstream robotics applications, including: (i)~efficient data generation and action prediction in imitation learning, (ii)~expressive dynamics and reward modeling in reinforcement learning, (iii)~scalable policy evaluation, and (iv)~visual planning for model predictive control.

%% file: sections/applications/app_imitation_learning.tex
\subsection{Cost-Effective Data Generation and Action Prediction in Imitation Learning}
\label{sec:app_data_gen_il}
In recent years, impressive research advances in robotics have been driven by imitation learning on large-scale expert demonstrations, circumventing the well-documented challenges associated with explicitly modeling dynamical interactions between robots and their environments. For example, SOTA vision-language-action (VLA) models~\citep{intelligence2025pi_, team2025gemini, lee2025molmoact, hancock2025actions} have demonstrated remarkable capabilities in generalist language-conditioned robot manipulation, exhibiting strong task and environment generalization and robust recovery behaviors in the presence of disturbances. Scaling these foundation models—both in parameter size and training data volume—has proven essential to realizing such performance leaps, mirroring the scaling paradigms that underscored the success of large language models. However, the immense time and labor costs associated with collecting large-scale, high-quality expert demonstrations pose a critical bottleneck to further advancement.
To address this challenge, recent work~\citep{xie2025human2robot, qiu2025lucibot, jang2025dreamgen, wang2025language, feng2025vidar, yang2025roboenvision} employs video models as cost-effective data generators of expert demonstrations, bypassing the overhead of human-supervised data collection. 
Concurrently, recent research has explored the use of video models as policy backbones in imitation learning, seeking to harness the synergies between dynamics prediction and policy learning to ground proposed robot actions.
\Cref{tab:app_video_data_generation} summarizes related methods.

\begin{table}[t]
    \centering
    \caption{Representative Video Models in Robot Data Generation and Policy Learning}
    \label{tab:app_video_data_generation}
    \begin{adjustbox}{width=\columnwidth}
    {
    \begin{tabular}{l l l c c l}
        \toprule
         Application & Method & Backbone & Architecture & Input Modality & Robot-Agnostic \\
         \midrule
         \multirow{4}{4cm}{IDM-Based Data Generation} & DreamGen~\citep{jang2025dreamgen} & Wan~2.1~\citep{wan2025wan} & DiT & I2V & \xmark\ (GR1, Franka, SO-100)\\
            & VPP~\cite{hu2024video} & SVD~\cite{blattmann2023stable} & U-Net & I2V & \xmark\ (OpenX) \\
            & ARDuP~\cite{huang2024ardup} & SD~\cite{rombach2022high} & U-Net & I2V & \xmark\ (Bridge)\\
            & Vidar~\cite{feng2025vidar} & Wan~2.2~\citep{wan2025wan} & DiT & I2V & \xmark\ (Aloha, AgiBot)\\
         \midrule
         \multirow{4}{4cm}{Motion Retargeting} & LuciBot~\cite{qiu2025lucibot} & KLING~\citep{KuaishouKlingAI2024} & DiT & I2V & \xmark\ (N/A)\\
            & AVDC~\cite{ko2023learning} & ADM-G~\citep{dhariwal2021diffusion} & U-Net & I2V & \xmark\ (Sawyer, Bridge, Franka)\\
            & VideoAgent~\cite{soni2024videoagent} & ADM-G~\citep{dhariwal2021diffusion} & U-Net & I2V & \xmark\ (Sawyer, Bridge)\\
            & Dreamitate~\cite{liang2024dreamitate} & SVD~\cite{blattmann2023stable} & U-Net & I2V & \xmark\ (xArm, UR5)\\
         \midrule
         \multirow{6}{4cm}{Policy Backbone} & UVA~\cite{li2025unified} & Scratch & DiT & I2V & \xmark\ (ARX X5)\\
            & UWM~\cite{zhu2025unified} & Scratch & DiT & I2V & \xmark\ (DROID)\\
            & UniVLA~\cite{wang2025unified} & Emu3~\citep{wang2024emu3} & DiT & I2V & \xmark\ (AgileX)\\
            & Video Policy~\cite{liang2025video} & SVD~\citep{blattmann2023stable} & U-Net & I2V & \xmark\ (DROID)\\
            & LVP~\cite{chen2025largevideoplannerenables} & Scratch & DiT & I2V & \xmark\ (N/A)\\
            & Mimic-Video~\cite{pai2025mimic} & Cosmos-Predict2~\cite{agarwal2025cosmos} & DiT & I2V & \xmark\ (Franka)\\
         \bottomrule
    \end{tabular}
    }
    \end{adjustbox}
\end{table}

\begin{figure}[t]
    \centering
    \includegraphics[width=0.85\linewidth]{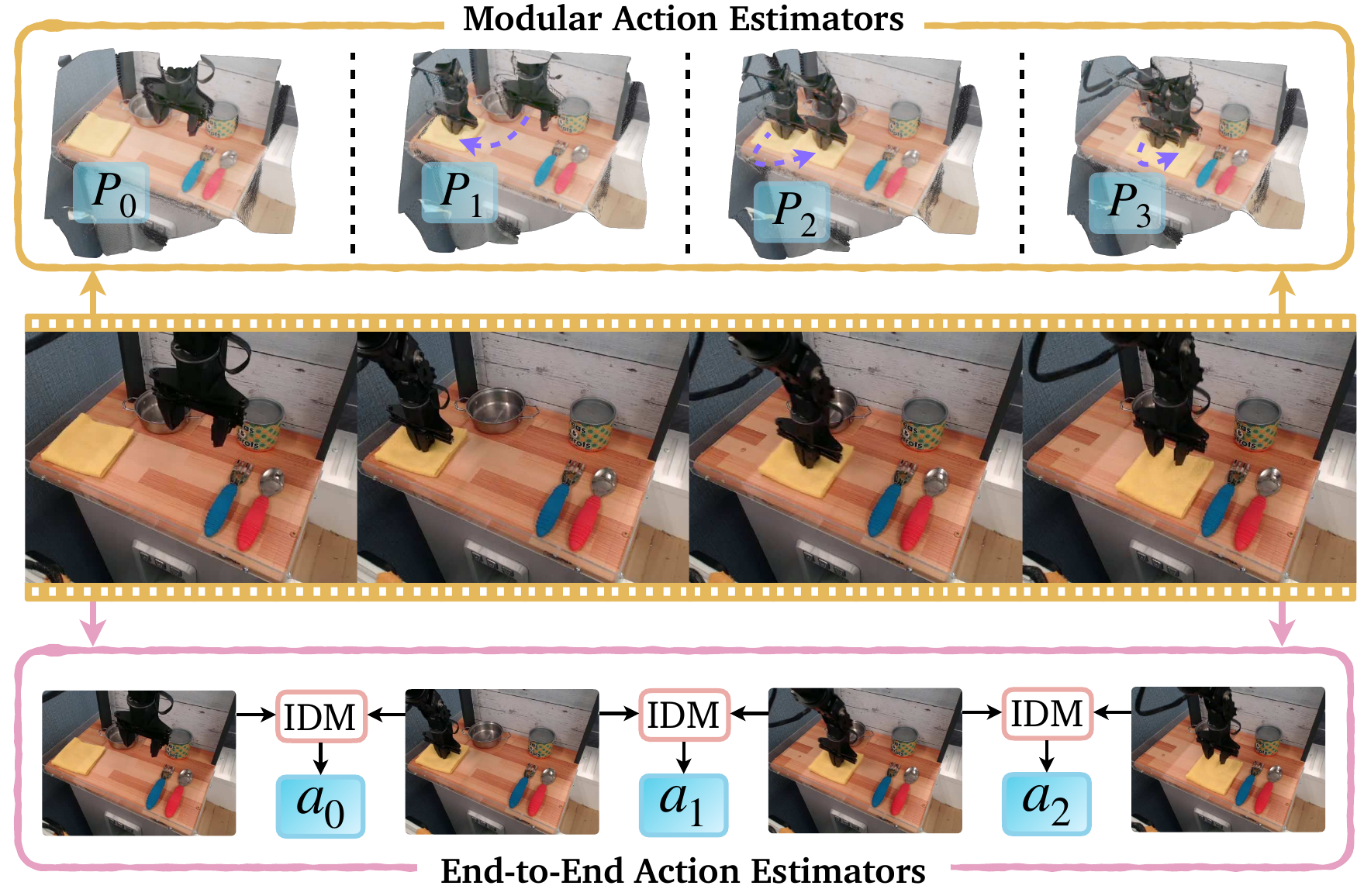}
    \caption{\textbf{Video Models for Data Generation.} Generative video models provide a scalable pathway for high-fidelity data generation, enabling cost-effective robot policy learning. Robot actions can be extracted from videos through modular approaches using end-effector pose tracking or end-to-end approaches, such as inverse-dynamics methods.
    }
    \label{fig:data_gen}
\end{figure}

\p{Video Models as Data Generators}
For scalable data generation, existing methods fine-tune pre-trained video models, such as Cosmos Predict \cite{NVIDIA_WorldSim_2025} and Wan \cite{wan2025wan}, on robot datasets for adaptation to robot embodiments and environments.
However, the expressiveness of text-conditioned and image-conditioned video models is often too limited to capture the diversity of the training data required to train generalizable robot policies. Subsequent research on controllable video models~\citep{wang2024motionctrl, zhang2025tora, fu2025learning, wang2025language} seek to overcome this limitation through finer control of camera and object motion in generated videos, given an input text, image, keypoints, or trajectory vectors, describing spatial changes in the scene temporally.
Some methods~\cite{yang2025roboenvision} further decompose the task into a set of subtasks for generating keyframes, which serve as conditioning inputs to generate videos for longer-horizon tasks.
Other methods~\citep{xie2025human2robot} train video models for effective cross-embodiment transfer, generating robot videos and extracting corresponding actions from models trained human video data. 
The resulting action-annotated data is used for policy learning, e.g., with VLA models \cite{jang2025dreamgen, NVIDIA_WorldSim_2025}, or directly applied to a robot \cite{ajay2023compositional, xu2025vilp, zhang2025gevrm}.
These methods typically recover robot actions from generated videos using an end-to-end approach or a modular approach. While modular approaches offer superior generalization, they are inherently prone to compounding errors across sequential stages, a limitation that end-to-end paradigms effectively circumvent.
\Cref{fig:data_gen} illustrates these approaches, contrasting the two distinct action prediction paradigms.

\emph{End-to-end} methods typically utilize latent action models (LAMs) or inverse-dynamics models (IDMs) to estimate actions from videos. LAMs~\cite{schmidtlearning, ye2024latent, chen2025villa} infer actions that explain the transition between a pair of video frames in latent space without the need for ground-truth actions, which often resides in a different action space. To learn latent actions without supervision, existing methods train an encoder-decoder model with a reconstruction-based loss function, typically implemented in latent space. The encoder takes in a pair of frames and outputs a latent action, which is passed into the decoder to recover the future frame. SOTA LAMs use VQ-VAEs as the model architecture for the encoder and decoder, given the superior tradeoff offered by VQ-VAEs in terms of expressiveness and training efficiency. Given the non-equivalence between the action space of the ground-truth and latent actions, LAMs often require a subsequent fine-tuning stage at deployment time to align the ground-truth and latent action spaces using a small set of action-labeled data.
Unlike LAMs, IDMs~\cite{agrawal2016learning, pathak2018zero, tan2025anypos} learn to predict actions from videos in a supervised fashion from action-labeled video data. Recent methods parameterize the video-to-action mapping using an encoder-decoder model trained with a diffusion objective. Although the need for ground-truth action labels poses a challenge to IDMs, IDMs do not require a fine-tuning stage, enabling zero-shot deployment in policy learning, either through direct action regression~\citep{jang2025dreamgen, hu2024video} or video-based active-region localization~\citep{huang2024ardup, feng2025vidar}.

Broadly, \emph{modular} approaches~\cite{patel2025robotic, qiu2025lucibot, bharadhwaj2024gen2act} estimate the target object's pose in each video frame, utilizing learned pose trackers~\cite{wen2024foundationpose, labbe2022megapose}, monocular depth estimators \cite{ke2025video}, optical flow predictors~\citep{ko2023learning}, or CAD models~\citep{liang2024dreamitate} for 3D object pose estimation. The resulting object trajectory is subsequently retargeted (i.e., applied) to the robot, under the assumption that a fixed rigid transformation exists between the object's reference point and the robot's end-effector. 
By taking advantage of low-level control routines, these methods can be deployed zero-shot without the need for a separate fine-tuning stage for aligning the action spaces between training and deployment.

\p{Video Models as Policy Backbones}
Unified video-action methods employ video models as policy backbones, co-training robot policies to jointly predict videos and actions conditioned on a language instruction and initial observation. While most unified video-action methods~\citep{radosavovic2023robot, cheang2024gr, li2025unified, zhu2025unified} utilize a VLA architecture for these policies, more recent work~\citep{liang2025video, chen2025largevideoplannerenables, pai2025mimic} directly adapts pre-trained video models for joint video and action generation using autoregressive transformer architectures~\cite{wu2023unleashing, cheang2024gr, radosavovic2023robot, li2025unified}, typically with diffusion-based supervision~\citep{guo2024prediction, liang2025video, zhu2025unified, zhang2025dreamvla, wang2025unified}. By jointly modeling future frame prediction and action estimation, these policies exploit the causal relationship between a robot's actions and subsequent environmental states, enabling effective inference-time optimization toward a desired goal. Furthermore, these unified architectures inherently support diverse downstream tasks, such as modeling forward and inverse kinematics. While pre-trained policy backbones instill robust physical priors that enhance out-of-distribution generalization, they can be challenging to fine-tune effectively compared to models trained from scratch.

%% file: sections/applications/app_rl.tex
\subsection{Dynamics and Rewards Modeling in Reinforcement Learning}
\label{sec:app_rl}

Although imitation learning provides a more sample-efficient training pipeline than reinforcement learning (RL), the resulting policies notoriously fail to generalize outside the training data distribution. RL mitigates these generalization bottlenecks; however, it requires well-defined dynamics models and reward functions, both of which are notoriously difficult to hand-engineer in complex physical domains. Further, due to its low sample efficiency, RL requires large amounts of training data to achieve the same level of performance as imitation-learned policies. To address these limitations, recent research explore video models as expressive backbones for dynamics and rewards modeling, as summarized in~\Cref{tab:app_dynamics_reward}

Whereas earlier works utilized recurrent state-space models~\cite{seo2022reinforcement} or image diffusion frameworks~\cite{jiang2025world4rl} for dynamics prediction, more recent approaches leverage video diffusion models, which yield higher-fidelity and more physically consistent rollouts. This training paradigm typically begins with a pre-training phase for text-conditioned video generation, followed by a post-training phase on action-aligned trajectories to achieve precise action-space alignment~\citep{hafner2025training}. Once aligned, these video world models can be directly deployed as dynamics and reward predictors during reinforcement learning (RL) policy fine-tuning~\citep{luo2024grounding}. Alternatively, reward signals can be derived by evaluating the generated video frames with vision-language models (VLMs)~\cite{xiao2025world}, or computed natively via video prediction likelihood~\citep{escontrela2023video} and conditional entropy metrics~\citep{huang2024diffusion}. Although VLMs offer superior generalization capabilities, they are susceptible to hallucinations that can introduce misleading signals, thereby impeding the learning process.

\begin{table}[t]
    \centering
    \caption{Representative Video Models in Dynamics and Rewards Modeling in Reinforcement Learning}
    \label{tab:app_dynamics_reward}
    \begin{adjustbox}{width=\columnwidth}
    {
    \begin{tabular}{l l l c c l}
        \toprule
         Application & Method & Backbone & Architecture & Input Modality & Robot-Agnostic \\
         \midrule
         \multirow{2}{4cm}{Dynamics Prediction} & World-Env~\cite{xiao2025world} & Scratch & U-Net & AC-I2V & \xmark\ (Franka) \\
            & GVM2A~\cite{luo2024grounding} & ADM-G~\citep{dhariwal2021diffusion} & U-Net & AC-I2V & \xmark\ (Franka, Sawyer) \\
         \midrule
         \multirow{2}{4cm}{Reward Prediction} & VIPER~\cite{escontrela2023video} & VideoGPT~\cite{yan2021videogptvideogenerationusing} & VQ-GAN & I2V & \xmark\ (RL-Bench) \\
            & Diffusion Reward~\cite{huang2024diffusion} & VQ-Diffusion~\cite{gu2022vector} & VQ-GAN & I2V & \xmark\ (Sawyer) \\
         \midrule
         Dynamics \& Reward Prediction & Dreamer~4~\cite{hafner2025training} & Scratch & DiT & AC-I2V & \xmark\ (Bridge) \\
         \bottomrule
    \end{tabular}
    }
    \end{adjustbox}
\end{table}

%% file: sections/applications/app_policy_eval.tex
\subsection{Scalable Policy Evaluation}
\label{sec:app_policy_eval}

\begin{table}[t]
    \centering
    \caption{Representative Video Models in Policy Evaluation}
    \label{tab:app_policy_eval}
    \begin{adjustbox}{width=\columnwidth}
    {
    \begin{tabular}{l l l c c l}
        \toprule
         Method & Backbone & Architecture & Input Modality & History-conditioned & Robot-Agnostic \\
         \midrule
         IRASim~\cite{zhu2024irasim} & Scratch & DiT & AC-I2V & \cmark & \xmark~(RT-1, Bridge) \\
         WorldEval~\cite{li2025worldeval} & Wan~2.1~\citep{wan2025wan} & DiT & AC-I2V & \xmark & \xmark~(AgileX) \\
         WorldGym~\cite{quevedo2025worldgymworldmodelenvironment} & Scratch & DiT & AC-I2V & \xmark & \xmark~(Bridge) \\
         Ctrl-World~\cite{guo2025ctrl} & SVD~\cite{blattmann2023stable} & U-Net & AC-I2V & \cmark & \cmark~(N/A) \\
         Cosmos-Eval~\cite{tseng2025scalable} & Cosmos-Predict2~\citep{agarwal2025cosmos} & DiT & AC-I2V & \cmark & \xmark~(Bridge) \\
         Veo-Robotics~\cite{veorobotics2025} & Veo~2~\citep{google2024veo2} & DiT & AC-I2V & \xmark & \xmark~(ALOHA~2) \\
         \bottomrule
    \end{tabular}
    }
    \end{adjustbox}
\end{table}

Beyond policy learning, video models offer a path towards reliable policy evaluation that is reproducible across different environments and task settings. Real-world evaluation of robot manipulation policies is notably challenging due to the significant inherent hardware and labor costs, especially for generalist robot policies, which typically require evaluation across a combinatorial number of operating environments. For example, each hardware trial requires human monitoring to reset the environment, observe the policy rollout, and record success scores, which is typically expensive.  
While hardware evaluation is the gold-standard, policy evaluation via physics-based simulation is the best available alternative, especially with recent improvements in visual fidelity, system identification, and careful tuning of material properties. However, physics-based simulation demands substantial initialization overhead, as each real-world environment must be manually reconstructed. This process requires the meticulous tuning of numerous parameters, such as asset lighting and material surface properties, to mitigate the sim-to-real gap.
In contrast, video models~\citep{li2024evaluating,zhang2025real,jain2025polaris} offer a higher-fidelity, more scalable framework for policy evaluation, modeling intricate robot-environment interactions that are challenging for physics-based simulators, e.g., in deformable-body simulation. 

We survey a growing body of research on methodologies for video-model-based policy evaluation, spanning diverse techniques for policy comparison and the estimation of real-world success rates, as summarized in~\Cref{tab:app_policy_eval}. These studies demonstrate that video world models can serve as high-fidelity verification tools prior to real-world deployment. Specifically, real-world success rates can be quantified by executing the policy in closed-loop simulations with a video model, conditioned on initial observation samples and task instructions~\cite{quevedo2025worldgymworldmodelenvironment,guo2025ctrl,huang2025enerverse,1x_world,li2025worldeval,zhu2024irasim,tseng2025scalable,veorobotics2025}.
To improve the quality and consistency of video generation, some architectures~\cite{li2025worldeval} condition the video model on the latent action representation in the policy, instead of conditioning on the robot's physical actions~\cite{guo2025ctrl,quevedo2025worldgymworldmodelenvironment}. 
Moreover, several architectures incorporate multi-view image conditioning to reduce hallucinations~\cite{guo2025ctrl}, effectively leveraging the multi-camera configurations typically deployed in robotic manipulation setups. To further mitigate prediction errors that accumulate over long rollouts, existing methods incorporate history during video generation, conditioning future video frames not only on the current observation but also on past observations. The history may consist of all frames within a fixed window length~\cite{zhu2024irasim,quevedo2025worldgymworldmodelenvironment,1x_world} or a sparsely sampled subset of past frames~\cite{guo2025ctrl,huang2025enerverse}. Moreover, augmenting observations with corresponding robot joint poses has been shown to improve frame-level action controllability~\cite{guo2025ctrl}. 

Each rollout is scored according to a rubric to quantify success (e.g., Bernoulli score, partial credit) on task completion and instruction following, and the scores of individual rollouts are aggregated into empirical success rates. 
To assess the accuracy of video models in policy evaluation, existing approaches~\cite{veorobotics2025,guo2025ctrl,tseng2025scalable} use metrics such as the Pearson correlation coefficient and the Mean Maximum Rank Violation (MMRV). The Pearson correlation coefficient measures the strength of the linear relationship between predicted and real-world success rates --- a high correlation is achieved when policies with high real-world success rates also exhibit high success rates in video model evaluations, and policies with low real success rates similarly have low video model success rates. In contrast, rank violation captures ranking inconsistencies between policy pairs, weighting each violation by the absolute difference in real-world success rates. The MMRV score is computed as the average across policies of their maximum rank violations. 
Although the predicted and real-world success rates do not always match, the evaluation results provide valuable insights for ranking the relative performance of policies, which is essential in improving robot policies in subsequent design and training stages.

Furthermore, video models can also be used to assess the robustness and safety of robot policies, particularly in out-of-distribution conditions~\cite{veorobotics2025,quevedo2025worldgymworldmodelenvironment}. Such out-of-distribution scenarios can be rapidly constructed via image editing~\cite{veorobotics2025}, e.g., to alter the background or add novel objects and distractors. This pipeline enables the use of video models to consistently predict performance degradation along different axes of generalization. Additionally, the pipeline can be readily interfaced with a VLM to generate safety-critical scenes and tasks for testing whether the policy executes semantically or physically unsafe behavior. 
Additionally, for accurate predictions across successful and failure scenarios, existing methods~\citep{1x_world,tseng2025scalable,zhu2024irasim} incorporate failure data when training video models to avoid a bias towards always predicting success, which we discuss further in~\Cref{subsec:chall_data_curation}.

%% file: sections/applications/app_planning.tex
\subsection{Visual Planning for Model Predictive Control}
\label{sec:app_visual_plan}

\begin{table}[t]
    \centering
    \caption{Representative Video Models in Visual Planning}
    \label{tab:app_visual_planning}
    \begin{adjustbox}{width=\columnwidth}
    {
    \begin{tabular}{l l l c c l}
        \toprule
         Method & Backbone & Architecture & Input Modality & History-Conditioned & Search  \\
         \midrule
         Visual~MPC~\cite{ebert2018visual} & Scratch & RNN & AC-I2V & \xmark & Sampling \\
         FLIP~\cite{gao2024flip} & Scratch & DiT & I2V & \cmark & Tree Search \\
         VLP~\cite{du2023videolanguageplanning} & Scrtach & U-Net & I2V & \xmark & Sampling \\
         MindJourney~\cite{yang2025mindjourney} & Wan~2.2~\citep{wan2025wan} & DiT & AC-I2V & \xmark & Tree Search \\
         SuSIE~\cite{black2023zero} & InstructPix2Pix~\citep{brooks2023instructpix2pix} & U-Net & I2I & \xmark & Greedy \\
         NovaFlow~\cite{li2025novaflow} & Wan~\citep{wan2025wan} & DiT & I2V & \xmark & Greedy \\
         \bottomrule
    \end{tabular}
    }
    \end{adjustbox}
\end{table}

Robot planning offers a compelling alternative to imitation learning, facilitating robust out-of-distribution generalization beyond the support of the training dataset. However, the design of high-fidelity models that closely predict the scene dynamics often proves prohibitively challenging. To address these drawbacks, recent work has explored utilizing video models in \textit{visual planning}, defined as the problem of synthesizing a sequence of images or video frames that show the steps necessary to complete a task specified by a language instruction and an initial observation. By leveraging the extensive diversity of the training data encoded by video models, these methods circumvent the need for large-scale expert demonstrations or explicit dynamics models to solve a broad range of robot tasks. 

Visual planning methods \cite{gao2025adaworld} generally optimize the generated video plans using sampling-based trajectory optimization methods, such as the gradient-free cross-entropy method, or gradient-based methods, e.g., gradient descent or Levenberg-Marquardt optimizers. The optimization routine is often embedded within a model predictive control (MPC) framework, facilitating the incorporation of new observations during planning via sensor feedback.
Video-model-based visual planning methods can utilize \textit{action-guided} or \textit{action-free} approaches to create feasible robot trajectories, as outlined in~\Cref{tab:app_visual_planning}.

\p{Action-Guided Visual Planning}
Action-guided visual planning methods utilize a three-step approach: first, generating action proposals; second, synthesizing video trajectories using video models as dynamics prediction modules conditioned on these proposals; and third, evaluating the resulting trajectories against a predefined objective function. Broadly, the action proposals are typically generated by sampling-based approaches~\cite{ebert2018visual}, learned approaches~\cite{gao2024flip}, or VLMs~\cite{yang2025mindjourney, du2023videolanguageplanning}. Sampling-based approaches~\cite{ebert2018visual} primarily employ the cross-entropy method with subsequent optimization of the generated frames based on photometric/Euclidean error signals with respect to user-provided goal images or keypoints. Other approaches extend this framework by generating action proposals via learned models, such as autoencoders~\cite{gao2024flip}, VLMs~\cite{du2023videolanguageplanning, yang2025mindjourney}, or auxiliary video models~\cite{cen2024using}, and subsequently scoring the candidate trajectories using either the proposal network itself or an external evaluation model~\cite{ma2023liv}.

\p{Action-Free Visual Planning}
In contrast, action-free visual planning methods bypass action proposal generation entirely, synthesizing future video plans directly from text-conditioned video models and utilizing the resulting frames as visual subgoals. For low-level execution, existing frameworks train goal-conditioned policies to derive robot actions from these image subgoals (e.g., via behavior cloning~\cite{black2023zero}) or employ IDMs to map current observations and subgoals directly to control actions~\cite{bu2024closed, du2023learning}. Alternatively, object pose trajectories can be extracted from the generated videos~\cite{li2025novaflow} and refined through least-squares optimization prior to motion retargeting.

%% file: sections/evaluation/evaluation.tex
\section{Evaluating Video Models}
\label{sec:eval_video_models}
In this section, we present standard metrics and benchmarks for evaluating video models across different dimensions, including perceptual quality, physical consistency, and semantic alignment with the input prompts.

\input{sections/evaluation/metrics}

\input{sections/evaluation/benchmarks}

\input{sections/evaluation/evaluation_axes}

%% file: sections/evaluation/metrics.tex
\subsection{Metrics for Evaluating Video Models}
\label{subsec:bkgd_metrics}

To assess video generation quality in accordance with human judgment, various metrics are needed to evaluate video models on visual quality, temporal coherence, diversity of generation, and physical commonsense. 
Although many traditional image quality metrics~\citep{Hore2010PSNR, wang2004image, fu2023dreamsimlearningnewdimensions} exist, these metrics do not assess temporal consistency of generated videos. To address this limitation, recent work~\citep{unterthiner2018towards, unterthiner2018towards} has introduced video-specific metrics that measure video quality both spatially and temporally. 
Additionally, application-focused metrics are equally important. For example, in policy evaluation, the usefulness of an action-conditioned video model depends on how well it can predict policy success rate, and in policy learning, whether the generated demonstration data results in higher-performance policies~\cite{jang2025dreamgen}.
Here, we discuss frame-level metrics and spatiotemporal consistency metrics.

\p{Frame-level Metrics} 
Traditional frame-level metrics from computer vision can be used to assess video generation quality, %
such as the peak signal-to-noise ratio (PSNR)~\citep{Hore2010PSNR} and structural similarity index (SSIM)~\citep{wang2004image}.
Unlike PSNR, which is inversely proportional to the mean squared error of the pixel values, SSIM assesses image quality on luminance, contrast, and structure, which is more indicative of perceived similarity. Pixel-based metrics often fail to capture higher-order image structures that align with human perceptual judgments~\cite{wang2004image}, which might be better represented in the learned feature spaces of deep models. For instance, the CLIP similarity score uses the cosine similarity between images in the embedding space~\cite{radford2021learning} to assess semantic alignment between a pair of images. For generative models, the inception score~\citep{salimans2016improved} assesses if the model confidently generates diverse, yet semantically meaningful images by computing the expected KL-divergence between the predicted class distribution and the marginal distribution of all generated images. However, the inception score does not use real samples, depends on a fixed label space, and is not suited for image generation without a target object. The Fr\'echet Inception Distance (FID)~\citep{heusel2018ganstrainedtimescaleupdate} addresses these challenges by computing the Wasserstein-2 distance between multi-dimensional Gaussian distributions over the real and generated feature embeddings. Likewise, the Learned Perceptual Image Patch Similarity (LPIPS)~\citep{zhang2018unreasonableeffectivenessdeepfeatures} is a similarity score on feature embeddings from several layers of a deep image recognition network, which has been shown to be comparable to human judgments. Furthermore, recent work~\citep{fu2023dreamsimlearningnewdimensions} has demonstrated that these perceptual metrics can be learned to further align with human judgments.

\p{Spatiotemporal Metrics}
While frame-level metrics focus on visual representation at any given video frame, assessing the temporal coherence of videos is also important for evaluating video generation quality. The Fr\'echet Video Distance (FVD)~\citep{unterthiner2018towards} accomplishes this by extending the Fr\'echet Inception Distance temporally to video representations that capture motion features across frames as well as the quality of images per frame. 
Instead of assuming a Gaussian representation for the features, the Kernel Video Distance (KVD)~\citep{unterthiner2018towards} uses a kernel-based metric (maximum mean discrepancy), to capture higher-order variations. The Fr\'echet Video Motion Distance (FVMD) computes the Fr\'echet distance on motion features by extracting keypoints that can be tracked across multiple frames and computing their velocities and accelerations. 
Prior work has also employed optical flow~\citep{manasa2016optical, wan2025wan, gibson1951perception} to assess temporal coherence and VLMs~\citep{bansal2024videophy, jang2025dreamgen} to evaluate physical consistency of generated videos.

%% file: sections/evaluation/benchmarks.tex
\subsection{Benchmarks for Evaluating Video Models}
\label{subsec:bkgd_benchmarks}
Despite their impressive capabilities, video models tend to generate videos that violate specific desired qualities, such as physical consistency, even when these videos are of high visual quality. Prior work~\cite{chi2024eva, li2025worldmodelbench, PAIBench2025} has introduced benchmarks to assess the performance of video models across different criteria, spanning  visual quality, dynamic consistency, and instruction following.
By evaluating video models across many dimensions, these benchmarks highlight critical areas for future research, in addition to identifying SOTA video models.
Generally, existing benchmarks demonstrate that video models fail to follow physical laws, even after scaling these models, although their aesthetic quality and temporal consistency tend to improve with scale.
While many existing benchmarks evaluate the overall quality or physical consistency of generated videos from video models, only a few benchmarks assess the safety of generated videos. These safety benchmarks~\citep{miao2024t2vsafetybenchevaluatingsafetytexttovideo, chen2024safewatchefficientsafetypolicyfollowing} provide an extensive evaluation of the safety of video models, demonstrating their tendency to generate illegal or unethical videos that violate their safety guidelines. 
In the subsequent discussion, we present benchmarks that examine the overall quality of video models broadly and those that focus on evaluating physical commonsense.

\p{Broad Benchmarks}
WorldModelBench~\citep{li2025worldmodelbench} introduces a benchmark to evaluate the instruction following and physics adherence capabilities of video models, identifying physically inconsistent changes in the sizes of objects, which violate the law of mass conservation.
Further, EvalCrafter~\citep{liu2024evalcrafter} evaluates the quality of generated videos in terms of their aesthetic and motion quality as well as temporal consistency and text-to-video alignment. While visual quality is assessed using DOVER~\citep{wu2023exploring}, motion quality is evaluated using action recognition methods, which are based on the activity classes in the Kinetics~400 dataset~\citep{kay2017kineticshumanactionvideo}. Temporal consistency and text-to-video alignment are measured using optical flow and CLIP, respectively.
EvalCrafter shows that its evaluation criteria are well-aligned with human preferences.
EWMBench~\citep{yue2025ewmbench} also evaluates video models along similar axes but uses the cosine-similarity metric with DINOv2~\citep{oquab2024dinov2learningrobustvisual} embeddings to measure visual consistency.
Furthermore, VBench~\citep{huang2024vbench} examines the performance of video models across sixteen fine-grained criteria, such as background consistency, subject consistency, and temporal flickering, in addition to aesthetic and semantic quality and instruction following. 
Like other benchmarks, VBench demonstrates that the results of each video model on each criterion is highly correlated with human preferences. 
Likewise, PAI-Bench~\citep{PAIBench2025} compares video models based on the temporal consistency, motion smoothness, and aesthetic quality of their generated videos.
Beyond measuring overall video consistency, T2V-CompBench~\citep{sun2025t2v} evaluates object attribute consistency, such as color, shape, and texture, along with action consistency in generated videos. Similarly, WorldSimBench~\citep{qin2024worldsimbench} assesses the physical consistency of motion, particularly the perception of a sense of depth, and changes in the velocity of objects in different media such as air and water.

\p{Physical Commonsense}
Other benchmarks focus primarily on evaluating the physical commonsense of video models~\citep{kang2024far, motamed2025generativevideomodelsunderstand}.
For example,  Physics-IQ~\citep{motamed2025generativevideomodelsunderstand} assesses video models' understanding of the laws of physics, such as those on optics, thermodynamics, magnetism and fluid dynamics, revealing that contemporary models lack a foundational grasp of physical laws.
Likewise, PhyGenBench~\citep{meng2024towards} measures the commensense knowledge of video models on $27$ physical laws, including gravity, sublimation, solubility, and friction, across $160$ prompts.  VideoPhy~\citep{bansal2024videophy} evaluates models along analogous axes, placing a stronger emphasis on complex object interactions.
Notably, $\text{VP}^{2}$~\citep{tian2023control} examines the alignment of video models with physical laws in the task of visual planning, demonstrating that strong performance on perceptual metrics is generally not indicative of alignment with physical consistency. Further, $\text{VP}^{2}$~\citep{tian2023control} shows that although scaling the model size and training dataset improves performance, the resulting gains plateau relatively quickly.

%% file: sections/evaluation/evaluation_axes.tex
\subsection{Horizon of Video Model Evaluation}
\label{subsec:eval_axes}
A unified framework for evaluating video models remains lacking, especially in the context of robotics applications. 
Existing metrics for evaluating video models generally assess either the perceptual quality~\citep{Hore2010PSNR, wang2004image, heusel2018ganstrainedtimescaleupdate} of their generated videos or their semantic consistency~\citep{radford2021learning, oquab2024dinov2learningrobustvisual}. While visual quality may be prioritized in content generation, physical consistency and predictive accuracy are more critical in robotics~\cite{tian2023control}. For example, robotics applications demand high-fidelity video generation to ensure consistency between predicted success rates and real-world success rates in policy evaluation and to mitigate compounding errors in downstream control applications.
Given the lack of suitable metrics, researchers often resort to surrogate measures based on the downstream application, e.g., correlation between  real and predicted policy success rates by video models, to approximate the performance of the video models~\cite{guo2025ctrl}.
Additionally, many works rely on human judgment to evaluate video model performance~\cite{wang2025language}, which is qualitative, costly, and subject to bias. 
Designing effective evaluation metrics is critical in assessing the performance of video models, especially along the axes of physical, temporal, and semantic consistency.

\p{Physical Consistency}
Physical consistency constitutes the most critical axis of evaluation for the successful integration of video models into robotics workflows. To serve as viable substitutes for physics-based simulators, generative video models must yield future predictions that adhere to fundamental physical laws. However, current architectures remain limited by non-causal dynamical artifacts, such as objects appearing or disappearing spontaneously, or morphing without external interaction or environmental triggers. Existing evaluation metrics fail to sufficiently capture this physical alignment due to their systemic bias toward surface-level perceptual fidelity. Furthermore, while VLMs can offer coarse feedback on consistency, they are prone to hallucinations that compromise evaluation reliability. Consequently, physical violations persist as a core vulnerability of SOTA video models, underscoring the urgent need for benchmarks that directly assess physically grounded video generation. Promising directions include video understanding methods that track cross-frame agent and object dynamics to systematically quantify physical anomalies, thereby providing the granular feedback necessary to guide the development of next-generation, physically consistent world models. This architectural progress will, in turn, accelerate the deployment of video models as high-fidelity data generators, robust predictors of policy success, and trustworthy visual planners.

\p{Temporal Consistency}
Maintaining temporal coherence remains a formidable challenge for video models, particularly during the long-horizon generation tasks crucial for robotic control. Although traditional visual fidelity metrics, such as SSIM and PSNR, offer coarse indicators of temporal consistency, their efficacy is strictly bounded to scenarios characterized by low per-frame pixel velocities. Conversely, robotics applications demand high-fidelity kinematic alignment across frames to support precise closed-loop control in visual planning and fine-grained policy rollouts. Current evaluation suites fail to probe temporal consistency at this granular level. For instance, existing frameworks typically measure temporal quality via optical flow or cross-frame feature similarity using learned encoders like DINO or CLIP, which operate at a semantic granularity too coarse to isolate fine-grained temporal artifacts. These limitations necessitate the design of more expressive benchmarks to systematically track and accelerate research progress.

\p{Semantic Consistency}
Video models occasionally generate content that is misaligned with the input prompt—a drawback that is particularly pronounced in image-and-text-conditioned generation. This semantic misalignment is highly consequential in data augmentation and visual planning, where the synthesized rollouts are treated as ground-truth data for training policies or for direct execution. Such discrepancies compromise safety and risk catastrophic real-world failures, underscoring the critical nature of this evaluation axis. Current verification methods primarily rely on CLIP to assess text-to-video semantic consistency; however, CLIP’s underlying contrastive framework effectively behaves as a bag-of-words model, rendering it insensitive to complex syntactic relationships and spatial compositionality in rich generation tasks. While VLMs offer finer-grained semantic probing, their propensity to hallucinate compromises their reliability in automated evaluation pipelines. Promising alternatives include task-understanding frameworks that explicitly verify the logical alignment between user-specified conditioning inputs and downstream video outputs.

%% file: sections/challenges_and_future_work/challenges_and_future_work.tex
\section{Open Challenges and Future Directions}
\label{sec:challenges_and_future_work}
We identify open research challenges in robotics applications of video models  and highlight directions for future research to address these challenges. We note that solutions to these challenges apply broadly beyond robotics, which could motivate novel applications of video models.

\input{sections/challenges_and_future_work/chall_hallucination_and_physics_viol}
\input{sections/challenges_and_future_work/chall_uq}
\input{sections/challenges_and_future_work/chall_instr_follow}
\input{sections/challenges_and_future_work/chall_evaluation}

\input{sections/challenges_and_future_work/chall_safe_content}
\input{sections/challenges_and_future_work/chall_safe_robot_interactions}
\input{sections/challenges_and_future_work/chall_action_estimation}

\input{sections/challenges_and_future_work/chall_long_horizon}

\input{sections/challenges_and_future_work/chall_data_curation}
\input{sections/challenges_and_future_work/chall_training}

%% file: sections/challenges_and_future_work/chall_hallucination_and_physics_viol.tex
\subsection{Hallucinations and Violations of Physics}
\label{subsec:chall_hallucinations_physics_violations}
Despite their impressive capabilities, video models often hallucinate, generating implausible video frames that are temporally inconsistent or misaligned with physical reality. 
In T2V generation, recent works~\cite{rawte2025vibe, chu2024sora, mei2025confident} have explored different types of hallucinations, including vanishing subject, omission error, numeric variability, visual incongruity, and subject dysmorphia~\cite{rawte2025vibe}, and have proposed hallucination detectors.
In robotics, the challenge of hallucination is especially important, given the critical role of accurate future prediction in policy evaluation and visual planning, among other application areas. 
Recent works have found that using multi-view frame inputs and particularly including a wrist camera view reduces hallucinations~\cite{guo2025ctrl}.
However, augmented conditioning inputs have limited effect on mitigating the violation of physics.
Specifically, video models typically fail to follow the laws of physics governing object motion and interactions~\cite{lin2025exploring, mei2025world}. Prior work~\cite{bansal2024videophy, motamed2025generativevideomodelsunderstand, li2025worldmodelbench} has demonstrated that generated videos from these models violate fundamental principles, such as Newton's law of motion, conservation of energy and mass, and gravitational effects, revealing a lack of understanding of these laws. Further, these models tend to mimic the closest training example when presented with a task at inference, limiting their generalization to new (unseen) tasks. In this setting, video models exhibit a propensity to prioritize transferring the color, size, velocity, and shape of objects in the training dataset to the objects in the new task, in that particular order~\cite{kang2024far, motamed2025generativevideomodelsunderstand}.

Further, existing work~\cite{bansal2024videophy, meng2024towards} shows that video models struggle with generating physically-realistic solid-solid interactions, demonstrating that video models do not understand the material properties of objects, the law of conservation of momentum, and the impenetrability of objects. Video models also show a fundamental lack of understanding of fluid mechanics and conservation of mass~\cite{motamed2025generativevideomodelsunderstand}, generating unrealistic videos of liquid flows, e.g., pouring a drink into a cup without a corresponding change in the volume of liquid in the cup.
However, generating physically realistic data is crucial in robotics, such as in robot learning where the generated data serves as expert demonstrations for training a policy, highlighting the need to impart physical understanding to video models.
Moreover, prompt engineering and scaling techniques do not adequately resolve this challenge, suggesting that novel architectures and training techniques are required to address this problem.
Next, we discuss potential directions for future research to improve the physical consistency of videos generated by video models.

\p{Future Directions}
Physical realism of generated videos from video models is essential in applications requiring trustworthy video generation, such as policy evaluation, planning, and policy learning, underscoring the importance of research on hallucination mitigation. To improve the physical consistency of generated videos, prior work~\citep{greydanus2019hamiltonian, allenblanchette2020lagnetviplagrangianneuralnetwork, liu2024physgen, li2025wonderplaydynamic3dscene} has explored integrating physics-based priors and physics-based simulation with video generation. 
Hamiltonian and Lagrangian approaches~\citep{greydanus2019hamiltonian, allenblanchette2020lagnetviplagrangianneuralnetwork} train models to predict the dynamics of system based on Hamiltonian or Lagrangian mechanics. By encoding the laws of physics as priors in the training process, the resulting models tend to adhere better to the physical laws. Other methods employ simulation engines to enforce physical laws.
For example, PhysGen~\citep{liu2024physgen} extracts object segmentation masks and physical properties using VLMs and computes feasible trajectories for the annotated objects using rigid-body dynamics equations. Subsequently, the trajectories are applied to corresponding pixels for video generation. However, the resulting video often contains artifacts, which degrades its fidelity. PhysGen uses a video diffusion model to edit these artifacts to generate higher-quality videos~\citep{meng2021sdedit}.
Likewise, WonderPlay~\citep{li2025wonderplaydynamic3dscene} constructs a 3D scene representation from the conditioning inputs and estimates the physical properties of objects in the scene using a VLM. The scene and objects' physical properties are passed into a physics-based simulator, which computes a coarse trajectory conditioned on applied forces.
The coarse trajectory from the simulator serve as conditioning inputs for a video model to generate future frames.
Other methods~\citep{zhang2025thinkdiffusellmsguidedphysicsaware} have utilized LLMs to refine video models' input prompts to provide comprehensive descriptions of physical attributes and interactions to improve the physical accuracy of generated videos.
Although these methods improve physical alignment, they do not eliminate violations of the laws of physics. Specifically, LLMs/VLMs are prone to hallucinations~\citep{liu2024survey}, which limits the effectiveness of the aforementioned methods.
Moreover, these methods rely on ad-hoc solutions, negatively impacting their generality and ease of implementation.
We believe a promising future direction will be to explore efficient techniques for natively endowing video models with a fundamental understanding of physical laws, which might necessitate the design of novel training techniques and model architectures~\citep{karniadakis2021physics} that enforce physical laws.

Additionally, improving the ability of video models to understand feasible interaction modes could be critical to mitigating hallucinations. Prior work has explored identifying and localizing functional interaction cues from videos, e.g., how humans interact with objects and environments, which is referred to as affordance-based video understanding~\citep{yu2023fine, li2022discovering}. 
Existing methods~\citep{fang2018demo2vec, nagarajan2019grounded, liu2022joint, murlabadia2023multi} localize interaction regions in videos and embed these cues into a latent space to predict the evolution of contact regions (hotspots) across future video frames.
The resulting affordance maps~\citep{bahl2023vrb, ju2024roboabc, li2025preciseaffordance, liu2020forecasting} can serve as conditioning (guidance) signals in video synthesis to improve the physical consistency of generated videos. Developing effective strategies for incorporating these signals during video generation is an important area for future work.

%% file: sections/challenges_and_future_work/chall_uq.tex
\subsection{Uncertainty Quantification}
\label{subsec:chall_uq}
Uncertainty quantification (UQ) techniques are widely utilized with traditional deep neural networks to examine the trustworthiness of these difficult-to-interpret models. While the rapid development of LLMs has led to increased research in UQ of large generative models to address severe hallucinations (see~\cite{shorinwa2025survey} for a review of LLM UQ methods), UQ methods for image and video generation models remain largely underexplored.
Meanwhile, extending existing UQ methods to video generation models faces significant challenges due to the complexity of spatial and temporal relationships. For example, generative video modeling fails to satisfy the central assumptions of standard Bayesian UQ methods, such as the assumption of independent, identically distributed samples, given the correlation between frames across multiple timesteps. Additionally, the significant computational cost associated with video generation impedes the application of ensemble-based UQ methods. Importantly, existing video models lack the ability to express or verbalize their confidence. As a result, inference-time UQ methods, e.g., black-box LLM UQ methods, cannot be directly applied to video models.

\p{Future Directions}
A few recent works have explored UQ techniques for controllable video generation, including text-conditioned and action-conditioned models.
S-QUBED~\citep{mei2025confident} quantifies uncertainty of T2V generation in the semantic space. However, this method only estimates task-level confidence on video generation.
On the other hand, $C^3$~\citep{mei2025world} trains video models for simultaneous video generation and uncertainty quantification in latent space. Specifically, $C^3$ enables video models to predict the uncertainty associated with each subpatch of the generated video, providing dense confidence estimates spatially and temporally.
In general, these methods are only guaranteed to provide calibrated uncertainty estimates within the distribution of the training dataset, limiting their effectiveness in out-of-distribution use-cases.
A promising direction for future work will be to explore more cost-effective methods for uncertainty quantification with provable guarantees both within and beyond the training distribution.

%% file: sections/challenges_and_future_work/chall_instr_follow.tex
\subsection{Instruction Following}
\label{subsec:chall_instr_follow}
Like early language models~\cite{ouyang2022training}, text-conditioned video models struggle with following user instructions specified in the input prompts~\cite{bansal2024videophy, qin2024worldsimbench}, which poses a significant limitation. Although these models are generally trained with guidance for alignment with the conditioning inputs, existing guidance mechanisms do not provide sufficient supervision to produce videos consistent with the input prompt. In general, SOTA video models often fail to complete the specified task due to their inability to extract and transfer the intended action from the conditioning input to the specified actor (agent). Prior work~\cite{li2025worldmodelbench} has shown that video models are generally able to correctly generate videos containing the agents stated in the input prompt but only partially follow the specified action and in some cases, completely fail to incorporate the specified action into the generated video.

Furthermore, video models typically fail to generate high-quality text (annotations) in videos~\cite{liu2024evalcrafter}, even in videos with otherwise high-fidelity components. This limitation becomes especially conspicuous when the input prompt specifically requests text annotations. Further, camera motion control via input prompts remains a major challenge for video models~\cite{yue2025ewmbench, liu2024evalcrafter}. Even when asked to generate videos with a static camera viewpoint without panning, video models tend to mimic their training videos, which often contain camera motion, resulting in non-adherence to the input prompt. 

These challenges limit robotics applications in data generation and policy learning, among others. For example, many robot data generation methods operate under the assumption of a static camera position for accurate $3$D pose estimation of the robot's end-effector pose through back-projection. Violation of this assumption results in inaccurate goal poses for tracking, degrading the robot's task performance. Likewise, failure to generate videos that correctly follow the actions specified by the input prompt leads to training data corruption, limiting the effectiveness of imitation learning, which relies on high-quality (expert) demonstrations.

\p{Future Directions} Recent work~\citep{fang2024vimi, xing2025aid, li2025trainingfreeguidancetexttovideogeneration} has explored improving the instruction-following capability of video models through multimodal conditioning. ViMi~\citep{fang2024vimi} interleaves language and images into a single instruction prompt, which is passed into a VLM to extract a conditional embedding for video generation. Similarly, Aid~\citep{xing2025aid} uses a VLM to predict the states of future video frames, which are fused with the text instruction to generate a conditional embedding for video synthesis. Other approaches~\citep{zhang2024interactivevideousercentriccontrollablevideo, wang2025atitrajectoryinstructioncontrollable} enable fine-grained multimodal user inputs to more strongly guide video models in the generation process. With InteractiveVideo~\citep{zhang2024interactivevideousercentriccontrollablevideo}, users can control the content of generated videos through image, text, and trajectory prompts, describing the desired motion of different elements of the video scene. Similarly, ATI~\citep{wang2025atitrajectoryinstructioncontrollable} enables localized control of deformations using keypoints and motion paths. Some other methods~\citep{yuan2024instructvideo} have examined instruction fine-tuning using preference-based reward models to train video models that are better at following user instructions. 
The aforementioned approaches generally rely on VLMs or other learned models that are susceptible to hallucinations, limiting their practical effectiveness.
To address this challenge, a promising research direction will be to explore intrinsic methods for improving task understanding and instruction following, e.g., through inference-time ``reasoning" over the generated video patches and frames, similar to the approach used in reasoning language models~\citep{guo2025deepseek, mei2025reasoninguncertaintyreasoningmodels}.

%% file: sections/challenges_and_future_work/chall_evaluation.tex
\subsection{Evaluating Video Models}
\label{subsec:chall_eval_video_models}
As established in \Cref{subsec:eval_axes}, robust evaluation metrics and standardized benchmarks for video world models remain notably deficient. To address these limitations, this section outlines several promising research directions aimed at bridging this methodological gap.

\p{Future Directions}
Prior works have developed benchmarks for evaluating video generation in terms of alignment with
downstream task performance~\cite{tian2023control, qin2024worldsimbench}. However, the environment complexity and visual quality of these benchmarks are fundamentally limited by the simulators. Recent works have explored benchmarking and evaluating T2V generation without ground-truth videos along many axes, such as overall video quality, text-to-video alignment, motion quality, and temporal consistency~\cite{liu2024evalcrafter}. To evaluate the quality of video models as embodied world models, WorldModelBench~\citep{li2025worldmodelbench} trains a VLM-based judge to assess video models in terms of instruction following, physics adherence, and commonsense.
In general, these benchmarks still lack the ability to assess video generation quality in fine-grained robot manipulation tasks. 
A promising direction for future research will be to explore robotics-centric benchmarks and multi-dimensional metrics for quantitative, efficient, and task-relevant video model evaluation. Further, future research on evaluation pipelines that compare the physical consistency of generated videos based on the corresponding motion in the associated $3$D scene reconstructions will be important.

%% file: sections/challenges_and_future_work/chall_safe_content.tex
\subsection{Safe Content Generation}
\label{subsec:chall_safe_content}
Many video models lack adequate safety guardrails, impeding their integration into many real-world applications.
Prior work~\citep{miao2024t2vsafetybenchevaluatingsafetytexttovideo, chen2024safewatchefficientsafetypolicyfollowing} has demonstrated the propensity of video models to generate unsafe content, containing crime, offensive activities, violence, or misinformation, which could hinder adoption in sensitive applications.
Despite the importance of safety, only a few papers~\citep{yoon2025safreetrainingfreeadaptiveguard} have explored methods for improving the safety of video models.
Addressing this challenge is essential to driving broader applications in robotics.

\p{Future Directions}
Enforcing the safety of video models is particularly challenging, given the broad diversity of unsafe video content. For tractability, prior work has utilized model guardrails to prevent the generation of potentially harmful content. While there is extensive literature on safeguarding LLMs~\citep{bai2022traininghelpfulharmlessassistant}, analogous techniques for video generation models are far less developed. Recent work (e.g., SAFEWatch~\citep{chen2024safewatchefficientsafetypolicyfollowing}) introduces a mechanism for enforcing user-specified safety policies during video generation. However, these methods are task-specific which limits their amenability to more general applications, underscoring the need for more versatile safety guardrails.
Additionally, there is a growing need for safety benchmarks for video models.
Existing benchmarks~\citep{ chen2024safewatchefficientsafetypolicyfollowing, miao2024t2vsafetybenchevaluatingsafetytexttovideo} evaluate video models on a limited range of criteria, e.g., crimes, hate content, privacy violations, and abusive content; however, a broader evaluation of safety is required for real-world use-cases in robotics. Developing more effective safety guardrails and introducing more comprehensive safety benchmarks will be critical to improving the overall safety of video models in video synthesis.

%% file: sections/challenges_and_future_work/chall_safe_robot_interactions.tex
\subsection{Safe Robot Interaction}
\label{subsec:chall_safe_robot_interactions}
Beyond safe video synthesis, robots must interact safely with other objects and agents in their environments. Safety in robotics can be broadly classified into two categories: (i)~physical safety, which involves avoiding all forms of collisions; and (ii)~semantic safety, which involves avoiding situations deemed potentially harmful by common-sense knowledge, such as tossing sharp objects at other agents. These forms of safety have been underexplored in robotics applications of video models. However, the success of many robotics tasks depends significantly on satisfying these safety constraints, e.g., in visual planning.

\p{Future Directions}
As embodied world models, video models enable robots to assess the safety of proposed actions without executing these actions in the real-world, which is critical to preserving the safety of robots, other agents, and their environments.
To detect safety violations, some existing methods~\citep{veorobotics2025} generate video predictions from robot action proposals using video models. 
Other prior work~\citep{hsu2024safetyfilter, nakamura2025generalizing} perform robot safety filtering directly in the latent space of world models to predict failures and preempt their occurrence. Likewise, existing methods~\citep{seo2025uncertaintyawarelatentsafetyfilters, agrawal2025anysafeadaptinglatentsafety} quantify uncertainty in the latent space to guard against unseen out-of-distribution (OOD) hazards and other user-specified constraint violations. While these works make notable steps toward safe robot control, they have primarily been limited to Markovian state-based world models. Extending these ideas to the spatiotemporal latent spaces of video world models remains a challenge, constituting a important direction for future research.
Moreover, generalization beyond the training distribution remains a core challenge in long-tail or safety-critical scenarios because world models are inherently limited by the distribution of their training data.

%% file: sections/challenges_and_future_work/chall_action_estimation.tex
\subsection{Action Estimation}
\label{subsec:chall_action_estimation}
SOTA imitation-learned robot policies require high-quality data~\cite{walke2023bridgedata, khazatsky2024droid, o2024open}, which is often expensive to collect in the real-world. Although video generation models can address this challenge, the video generated by these models typically do not contain action-labeled frames, which is essential in learning robot policies. 
Some methods have explored estimating robot actions from videos; however, these methods~\cite{ye2024latent} typically fail to achieve the high level of accuracy required in imitation learning in fine-grained tasks, hindering the effectiveness of current solutions that integrate video models in policy learning frameworks. We discuss a few limitations of latent action models and inverse-dynamics models, the two main approaches for action estimation from robot videos.

Latent action models~\cite{ye2024latent, chen2025villa} estimate robot actions between a pair of video frames in a (discrete) latent space, defined by a fixed set of action codes (primitives). The expressiveness of the learned latent actions is strongly influenced by the size of the latent codebook. Scaling the size of the latent codebook is challenging in practice due to the significant training instabilities and higher compute cost associated with training latent action models with larger codebooks. Further, latent actions are generally difficult to interpret, which makes data curation and analysis more challenging. Additional real-world data is also required to fine-tune latent action models to predict actions that are compatible with physical robots.
Inverse-dynamics models~\cite{pathak2018zero, tan2025anypos} suffer from similar challenges and require lots of training data to sufficiently cover the space of robot actions. Like other imitation-learned models, inverse-dynamics models exhibit limited generalization outside of the training distribution, hindering real-world applications. 

\p{Future Directions}
For better interpretability, recent work~\citep{deepmind2025genie3} utilizes small latent codebooks, which facilitate mapping each latent action to interpretable conditioning inputs. However, this approach fails to scale to more complex robotics tasks which require larger codebooks to effectively learn latent actions.
Exploring model architectures that induce interpretable latent actions, without compromising the expressiveness of the latent action space, will be critical to advancing the adoption of video models in robot policy learning. Likewise, the development of robust training procedures for latent action models will be essential to their generalization in robotics.
Additionally, semi-supervised training techniques could enable efficient training of generalizable inverse-dynamics models with only a few human annotations, constituting a promising area for future work.

%% file: sections/challenges_and_future_work/chall_long_horizon.tex
\subsection{Long-Horizon Video Generation}
\label{subsec:long_video_generation}
To serve as effective world models in robotics tasks, video models must predict sufficiently long future horizons that match the duration of robotic tasks, which are often minutes long.
However, SOTA video models are limited to generating videos of only a few seconds in duration. For example, Veo~3.1~\citep{deepmind_veo3_techreport} generates 8-second long videos while Wan~2.5~\citep{wan2025wan} generates 10-second long videos. These durations are not long enough for informed decision-making in many robotics problems, such as visual planning. Current video generation pipelines require extending multiple short clips to create longer-duration videos; however, this approach often introduces artifacts that degrade the temporal coherence and physical consistency of the resulting videos. 
While SOTA video models excel in short-duration video generation tasks, scaling these models to longer horizons for robotics tasks remains an open challenge.

\p{Future Directions}
Several architectures attempt to address video consistency over long horizons. Broadly, these methods use frame compression, sampling schedule optimizations, or hierarchical frameworks to enable long video generation. For example, MALT~\citep{yu2025malt} encodes past segments into a compact latent memory vector to facilitate autoregressive generation; however, this approach remains susceptible to error accumulation~\citep{wang2025error}. FramePack~\citep{zhang2025packing} mitigates drift by compressing frame contexts based on importance and establishing early endpoints to anchor the generation process. 
TTTVideo~\citep{dalal2025one} and LaCT~\citep{zhang2025test} utilize test-time training (TTT) to dynamically encode history into model weights or neural hidden states during inference. However, the quality of the generated videos degrades as video length increases.
To improve video fidelity, Long-Context Tuning (LCT)~\citep{guo2025long} expands the context window to maintain dense attention across multi-shot scenes but is limited by the quadratic cost of self-attention, which imposes a computational ceiling on the maximum generation length.  In contrast to these approaches, Mixture of Contexts (MoC)~\citep{cai2025mixture} circumvents compressing history entirely by recasting generation as an information retrieval task via a sparse attention routing mechanism.
Diffusion Forcing~\cite{NEURIPS2024_2aee1c41} offers an alternative solution by training models to denoise tokens with independent noise levels, enabling variable-horizon generation that empirically improves stability in policy rollouts. On the other hand, NUWA-XL~\citep{yin2023nuwa} utilizes a hierarchical ``diffusion-over-diffusion" structure, where a global model generates sparse keyframes and local models recursively fill the intermediate gaps.
Despite these improvements, existing methods still struggle to generate minutes-long videos, impeding the their effectiveness in real-world applications. A promising direction for future work will be to design efficient techniques for extending the context window of video models without a prohibitive increase in computation cost for long-video generation.

%% file: sections/challenges_and_future_work/chall_data_curation.tex
\subsection{Data Curation Costs}
\label{subsec:chall_data_curation}
High-quality data is essential to training video models that are capable of synthesizing high-fidelity, physically-consistent videos across diverse tasks. Beyond video quality, diversity of the training data strongly influences the fidelity of generated videos~\cite{peng2025open, yang2024cogvideox}, especially for text-conditioned or action-conditioned video models, which require extensive data coverage for desirable results.
Although large volumes of videos exist on the internet, many of these videos lack good visual quality and descriptive text annotations, posing a challenge.
Specifically, many existing datasets, such as WebVideo-10M~\cite{bain2021frozen} and Panda-70M~\cite{chen2024panda}, have focused on scale rather than quality, aggregating tens of millions of videos along with their captions. As a result, these datasets suffer from inaccurate, non-descriptive video captions, blurry videos, and rapid shot changes between temporally-inconsistent clips, which negatively impact training.
SOTA methods~\cite{peng2025open, agarwal2025cosmos, yang2024cogvideox, wan2025wan} typically rely on expensive data preprocessing pipelines to identify good-quality video data for training. This pipeline consists of three broad stages: (i)~video splitting, (ii)~video filtering, and (iii)~video annotation. At the video splitting stage, candidate video data is temporally segmented into short continuous clips using classical shot detection tools. Videos that are too short are removed at this stage. The resulting clips are then processed by video filters using metrics that assess the visual quality, text quality, motion smoothness and jitter, among others, often with learned models.
Subsequently, many methods~\cite{peng2025open, agarwal2025cosmos} utilize VLMs~\cite{zhang2024video, lin2024vila, chen2024panda} for annotating the processed videos, pairing text descriptions with the video data for supervising video models. The VLMs are typically fine-tuned for higher-quality video captioning.
This pipeline typically requires human supervision, increasing data curation costs.

\p{Future Directions}
Recent video datasets, e.g., VidGen-1M~\cite{tan2024vidgen} and OpenVid-1M~\cite{nan2024openvid}, have explored different pre-processing techniques to filter videos using well-defined quality scores before training. These quality scores evaluate the aesthetics, temporal consistency, motion fidelity, and caption descriptiveness, identifying video clips that are more likely to be natural without unrealistic motion. Although these datasets generally provide higher-quality training data, their relatively small size impedes strong zero-shot generalization. 
Consequently, many SOTA video models~\cite{wan2025wan, agarwal2025cosmos, deepmind_veo3_techreport} utilize proprietary curated datasets with dense annotations by VLMs and humans to ensure high quality control.
However, VLMs are prone to hallucinations, and human-annotated data collection is expensive, posing significant challenges in high-fidelity video data collection. Exploring strategies for grounding VLMs to minimize the risk of hallucinations is a promising area for future work. Likewise, developing novel-view synthesis techniques to efficiently scale data collection to new scenes from a small set of high-quality videos will be essential to reducing the costs of data curation. Further, more accurate methods for splitting and filtering videos will be critical to curating highly diverse datasets with good temporal and spatial consistency, e.g.,~\cite{wang2025koala}.
In robotics, high-fidelity future prediction in both successful and unsuccessful task rollout or demonstrations is essential for video models to serve as good proxies for real-world environments. Notably, failure demonstrations are important to train video models that faithfully execute actions output by the policy. Recent studies suggest that incorporating failure data from autonomous policy rollouts is critical for training world models that exhibit action controllability~\cite{1x_world,tseng2025scalable,zhu2024irasim}. Without such data, generated videos can exhibit an optimistic bias toward success, e.g., via hallucinations that reposition objects for easier grasping or over-estimate grasp feasibility, which could result in real-world failures~\cite{1x_world}.

%% file: sections/challenges_and_future_work/chall_training.tex
\subsection{Training and Inference Costs}
\label{subsec:chall_training_overhead}
SOTA video models require large amounts of compute resources for training and inference.
Although the true cost of training video models is often confidential especially for closed-source models, the most cost-effective SOTA open-source video models require hundreds of thousands of dollars to train---e.g., \$200k for Open-Sora~2.0~\cite{peng2025open}.
Video models typically have billions of parameters, which is a major contributing factor to their significant training costs. 
Although latent diffusion~\citep{blattmann2023align} reduces the number of parameters required by these models, the corresponding reduction in training costs is often not sufficient. As a result, research on training video models has broadly remained limited to large research groups with deep budgets. Moreover, recent developments such as classifier-free guidance~\cite{ho2022classifier, dhariwal2021diffusion} for fine-grained input-conditioning introduce additional computational overhead. Reducing training and inference costs is critical to advancing broader applications of video models in robotics.

Additionally, high-fidelity video models~\citep{wan2025wan, agarwal2025cosmos, brooks2024video, deepmind_veo3_techreport} are often limited by significantly slow inference speeds.
For example, Veo 3~\cite{deepmind_veo3_techreport} generates about $12$ video frames per second on an NVIDIA A100 GPU. Although sufficient in some use cases, the slow inference time presents an important challenge in many robotics applications, such as visual planning, where closed-loop execution requires real-time feedback to the planner for robustness. 
Existing video model planners often take a few seconds to generate feasible action trajectories for a single episode~\cite{ko2025implicitstateestimationvideo}, which is not fast enough for real-time operation.

\p{Future Directions}
For faster training and inference, recent work~\citep{peng2025open, hafner2025training} has explored spatial and temporal compression techniques to reduce the number of costly attention operations during video generation.
Dreamer~4~\citep{hafner2025training} applies temporal attention sparsely to every fourth video frame by decoupling spatial and temporal attention. Similarly, OpenSora uses deep compression autoencoders~\citep{chen2025deepcompressionautoencoderefficient} to downsample input tokens at greater ratios to speed up inference by an order of magnitude. Likewise, Wan~\citep{wan2025wan} employs a feature cache mechanism to enable chunk-based video synthesis while preserving temporal continuity.
Some other methods utilize shortcut models~\cite{frans2024one} for finer control over the number of sampling steps required for video generation without any significant degradation in video quality.
Like shortcut models, consistency models~\citep{song2023consistency, feng2024matrix} enable efficient video diffusion using a single-step denoising process. 
Other video models utilize more classical speed optimization techniques for faster inference, e.g., quantization~\citep{zhao2025viditqefficientaccuratequantization} and model distillation~\citep{yin2025slow}.
Further research on these topics is required to improve inference speeds for real-time applications.

%% file: sections/back_matter/conclusion.tex
\section{Conclusion}
\label{sec:conclusion}
This survey reviews video models and their applications as embodied world models in robotics, identifying prevailing model architectures, conditioning modalities, and key capabilities of video models. Specifically, we categorize existing robotics applications into four broad classes: robot data generation and action prediction in imitation learning, dynamics and rewards modeling in reinforcement learning, policy evaluation, and visual planning.
We emphasize that these applications are underpinned by the remarkable ability of video models to learn fine-grained spatiotemporal relationships that govern the evolution of the state of real-world environments, which is essential for physically consistent future predictions.
Furthermore, we identify critical open research challenges and propose directions for future research, seeking to motivate broader applications of video models.

%% file: sections/back_matter/acknowledgments.tex
\section*{Acknowledgments}
Asher J. Hancock was supported by the National Science Foundation Graduate Research Fellowship Program under Grant No. DGE-2146755. Apurva Badithela is supported by the Presidential Postdoctoral Research Fellowship at Princeton University. The authors were partially supported by the NSF CAREER Award \#2044149, \#2107048, the Office of Naval Research (N00014-23-1-2148), the Sloan Fellowship, and Apple Inc. Any views, opinions, findings, and conclusions or recommendations expressed in this material are those of the author(s) and should not be interpreted as reflecting the views, policies or position, either expressed or implied, of Apple Inc.